\documentclass[acmtog,authorversion]{acmart}

\usepackage{bm}
\usepackage{bbm}

\usepackage{graphicx}
\usepackage{tabularx}
\usepackage{subcaption}
\usepackage{multirow}
\usepackage{amsmath}
\usepackage{mathrsfs}

\newcommand{\vect}[1]{\bm{#1}}
\newcommand{\tildevect}[1]{\vect{\tilde{#1}}}

\newcommand{\eqword}[1]{{\text{#1}}}

\usepackage{xspace}

\newcommand{\keep}[1]{}
\newcommand{\old}[1]{}

\DeclareMathOperator*{\argmin}{arg\,min}

\DeclareMathOperator*{\simloss}{sim}
\DeclareMathOperator*{\stopgrad}{sg}
\DeclareMathOperator*{\softmax}{softmax}
\newcommand{\norm}[1]{\lVert #1 \rVert}

\usepackage{url}
\usepackage{hyperref}
\newcommand{\fig}{Figure{}~}
\newcommand{\eqn}{Equation{}~}

\AtBeginDocument{%
  \providecommand\BibTeX{{%
    \normalfont B\kern-0.5em{\scshape i\kern-0.25em b}\kern-0.8em\TeX}}}

\setcopyright{acmcopyright}
\acmJournal{TOG}
\acmYear{2022}
\acmVolume{41}
\acmNumber{6}
\acmArticle{209}
\acmMonth{12}
\acmDOI{10.1145/3550454.3555435}

\acmSubmissionID{146}

\begin{document}

\title{Rhythmic Gesticulator: Rhythm-Aware Co-Speech Gesture Synthesis with Hierarchical Neural Embeddings}

%%
%% Authors.
\author{Tenglong Ao}
\email{aubrey.tenglong.ao@gmail.com}
\affiliation{
  \institution{Peking University}
  \streetaddress{No.5 Yiheyuan Road, Haidian District}
  \city{Beijing}
  \state{Beijing}
  \country{China}
  \postcode{100871}
}

\author{Qingzhe Gao}
\email{gaoqingzhe97@gmail.com}
\affiliation{
  \institution{Shandong University and Peking University}
  \country{China}
}

\author{Yuke Lou}
\email{louyuke@pku.edu.cn}
\affiliation{
  \institution{Peking University}
  \streetaddress{No.5 Yiheyuan Road, Haidian District}
  \city{Beijing}
  \state{Beijing}
  \country{China}
  \postcode{100871}
}

\author{Baoquan Chen}
\email{baoquan@pku.edu.cn}
\affiliation{
  \institution{SIST \& KLMP (MOE), Peking University}
  \streetaddress{No.5 Yiheyuan Road, Haidian District}
  \city{Beijing}
  \state{Beijing}
  \country{China}
  \postcode{100871}
}

\author{Libin Liu}
\authornote{corresponding author}
\email{libin.liu@pku.edu.cn}
\affiliation{
  \institution{SIST \& KLMP (MOE), Peking University}
  \streetaddress{No.5 Yiheyuan Road, Haidian District}
  \city{Beijing}
  \state{Beijing}
  \country{China}
  \postcode{100871}
}

%%
%% This command defines the author string for running heads.
\renewcommand{\shortauthors}{Ao, Gao, Lou, Chen, and Liu}

\begin{abstract}
    Automatic synthesis of realistic co-speech gestures is an increasingly important yet challenging task in artificial embodied agent creation. Previous systems mainly focus on generating gestures in an end-to-end manner, which leads to difficulties in mining the clear rhythm and semantics due to the complex yet subtle harmony between speech and gestures. We present a novel co-speech gesture synthesis method that achieves convincing results both on the rhythm and semantics. For the rhythm, our system contains a robust rhythm-based segmentation pipeline to ensure the temporal coherence between the vocalization and gestures explicitly. For the gesture semantics, we devise a mechanism to effectively disentangle both low- and high-level neural embeddings of speech and motion based on linguistic theory. The high-level embedding corresponds to semantics, while the low-level embedding relates to subtle variations. Lastly, we build correspondence between the hierarchical embeddings of the speech and the motion, resulting in rhythm- and semantics-aware gesture synthesis. Evaluations with existing objective metrics, a newly proposed rhythmic metric, and human feedback show that our method outperforms state-of-the-art systems by a clear margin.
\end{abstract}

%%
%% Article info.



\begin{teaserfigure}
  \includegraphics[width=\textwidth]{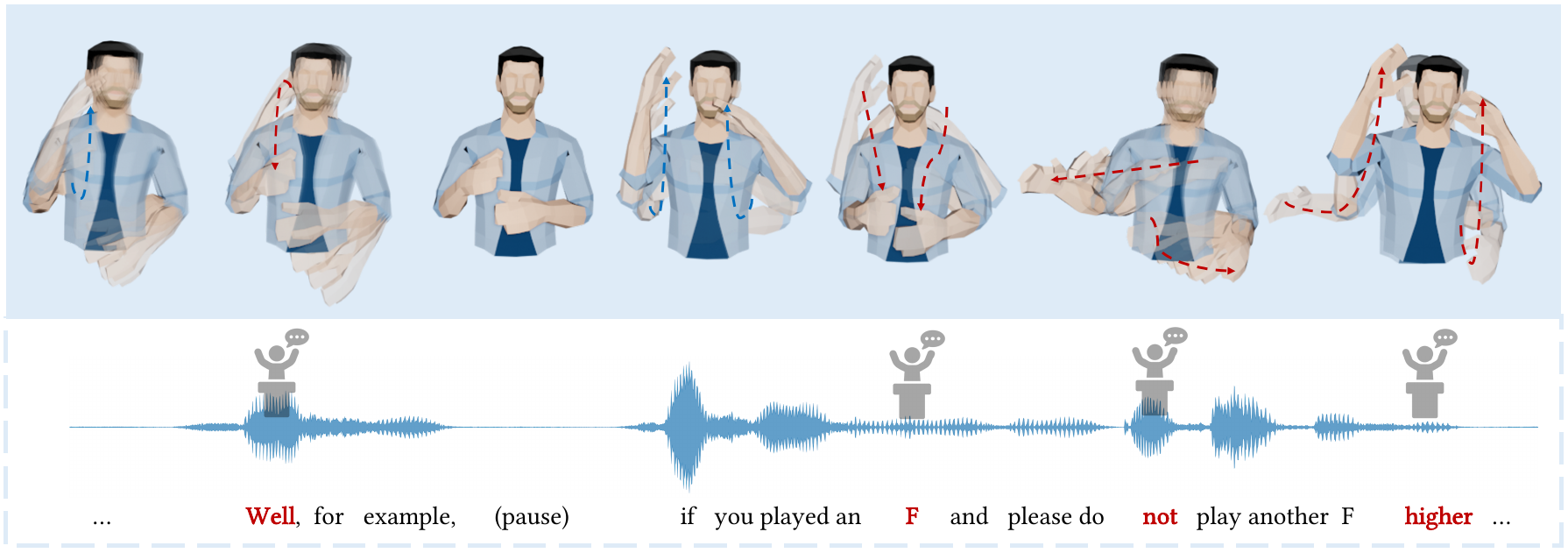}
  \caption{Gesture results automatically synthesized by our system for a beat-rich TED talk clip. The red words represent beats, and the red arrows indicate the movements of corresponding beat gestures.}
  \label{fig:teaser}
\end{teaserfigure}

\maketitle

%%
%% Sections.
\section{Introduction}
\label{sec:introduction}
Gesturing is an important part of speaking. It adds emphasis and clarity to a speech and conveys essential non-verbal information that makes the speech lively and persuasive \cite{burgoon1990nonverbal}. There are rich demands for high-quality 3D gesture animation in many industries, such as games, films, and digital humans. However, the difficulties in reproducing the complex yet subtle harmony between vocalization and body movement make synthesizing natural-looking co-speech gestures remain a long-standing and challenging task.

Gestures are grouped into six categories by linguists \cite{ekman1969repertoire, mcneill1992hand}---adaptors, emblems, deictics, iconics, metaphorics, and beats. Among them, the beat gestures are rhythmic movements that bear no apparent relation to speech semantics \cite{Kipp2004_Gesture} but serve meta-narrative functions~\cite{mcneill1992hand} that are crucial to rhythmic harmony between speech and gestures. Generating realistic beat gestures requires modelling the relation between the gestural beats and the verbal stresses. However, it has been observed that these two modalities are not synchronized in a strict rhythmic sense \cite{mcclave1994gestural}, making it difficult to learn their temporal connection directly from data using an end-to-end method \cite{kucherenko2020gesticulator, yoon2020speech, bhattacharya2021speech2affectivegestures}. 

Gestures are associated with different levels of speech information~\cite{mcneill1992hand}. For example, an emblem gesture such as \emph{thumbs-up} usually accompanies high-level semantics like \emph{good} or \emph{great}, while a beat gesture commonly comes with low-level acoustic emphasis. Many previous studies use only the features extracted at the last layer of an audio encoder to synthesize gestures \cite{kucherenko2020gesticulator, yoon2020speech, bhattacharya2021speech2affectivegestures, qian2021speech, alexanderson2020style}. This setup, however, may in effect encourage the encoder to mix the speech information from multiple levels into the same feature, causing ambiguity and increasing the difficulty in mining clear rhythmic and semantic cues.

In this paper, we focus on generating co-speech upper-body gestures that can accompany a broad range of speech content---from a single sentence to a public speech, aiming at achieving convincing results both on the rhythm and semantics. Our first observation is that gesturing can be considered as a special form of dancing under changing beats. We develop a rhythm-based canonicalization and generation framework to deal with the challenge of generating synchronized gestures to the speech, which segments the speech into short clips at audio beats, normalizes these clips into canonical blocks of the same length, generates gestures for every block, and aligns the generated motion to the rhythm of the speech. This framework, which is partially inspired by recent research in dance generation~\cite{aristidou2021rhythm}, provides the gesture model with an explicit hint of the rhythm, allowing the model to learn the pattern of gestural beats within a rhythmic block efficiently. Both the quantitative evaluation with a novel rhythmic metric and the qualitative evaluation with user studies show that the gestures generated by this pipeline exhibit natural synchronization to the speech.

As indicated in linguistics literature~\cite{Kipp2004_Gesture,Neff2008Gesture,Webb1996_Linguistic}, gestures used in everyday conversation can be broken down into a limited number of semantic units with different motion variations. We assume that these semantic units, usually referred to as \emph{lexeme}s, relate to the high-level features of speech audio, while the motion variations are determined by the low-level audio features. We thus disentangle high- and low-level features from different layers of an audio encoder and learn the mappings between them and the gesture lexemes and the motion variations, respectively. Experiments demonstrate that this mechanism successfully disentangles multi-level features of both the speech and motion and synthesizes semantics-matching and stylized gestures.

In summary, our main contributions in this paper are:
\begin{itemize}
    \item We present a novel rhythm- and semantics-aware co-speech gesture synthesis system that generates natural-looking gestures. To the best of our knowledge, this is the first neural system that explicitly models both the rhythmic and semantic relations between speech and gestures.
    \item We develop a robust rhythm-based segmentation pipeline to ensure the temporal coherence between speech and gestures, which we find is crucial to achieving rhythmic gestures.
    \item We devise an effective mechanism to relate the disentangled multi-level features of both speech and motion, which enables generating gestures with convincing semantics.
\end{itemize}
\section{Related Work}
\label{sec:related_work}

\subsection{Data-driven Human Motion Synthesis}
Traditional human motion synthesis frameworks often rely on concatenative approaches such as \emph{motion graph} \cite{Kovar2002_MotionGraphs}. Recently, learning-based methods with neural networks have been widely applied to this area to generate high-quality and interactive motions, using models ranging from feed-forward network \cite{holden2017phase,Starke2022_DeepPhase} to dedicated generative models~\cite{henter2020moglow, Ling2020_MotionVAE}. Dealing with the one-to-many issue where a variety of motions can correspond to the same input or control signal is often a challenge for these learning-based approaches. Previous systems often employ additional conditions, such as contacts \cite{starke2020local} or phase indices \cite{holden2017phase,Starke2022_DeepPhase}, to deal with this problem. Closer to the gesture domain is the speech-driven head motion synthesis, where conditional GANs \cite{sadoughi2018novel}, and conditional VAEs \cite{greenwood2017predicting} have been used.

\subsubsection{Music-driven Dance Synthesis}
Among the general motion synthesis tasks, music-driven dance generation addresses a similar problem to the co-speech gesture synthesis, where the complex temporal relation between two different modalities needs to be modeled accurately. Both motion graph-based methods \cite{kim2006making,chen2021choreomaster} and learning-based approaches \cite{Li_2021_aist,valle2021transflower,li2022bailando} have been adopted and successfully achieved impressive generation results. To deal with the synchronization between the dance and music, \citet{chen2021choreomaster} develop a manually labeled rhythm signature to represent beat patterns and ensures the rhythm signatures of the generated dance match the music. \citet{aristidou2021rhythm} segment the dance into blocks at music onsets, convert each block into a motion motif \cite{aristidou2018deep} that defines a specific cluster of motions, and use the motion motif to guide the synthesis of dance at the block level. \citet{li2022bailando} employ a reinforcement learning scheme to improve the rhythmic performance of the generator using a reward function encouraging beat alignment. Our rhythm-based segmentation and canonicalization framework is partially inspired by \cite{aristidou2021rhythm}. Similar to \cite{aristidou2021rhythm}, we also segment the gestures into clips at audio beats but learn a high-level representation for each clip via the vector quantization scheme \cite{oord2017neural} instead of the K-means clustering. Moreover, our framework generates gestures in blocks of motion and denormalizes the generated motion blocks to match the rhythm of the speech. In contrast, \citet{aristidou2021rhythm} synthesize dance sequences in frames conditioned on the corresponding motion motifs.

\subsection{Co-speech Gesture Synthesis}
The most primitive approach used to generate human non-verbal behaviors is to animate an artificial agent using the retargeted motion capture data. This kind of approach is widely used in commercial systems (e.g., films and games) because of its high-quality motion performance. However, it is not suitable for creating interactive content that cannot be prepared beforehand. Generating co-speech gestures according to an arbitrary input has been a long-standing research topic. Previous studies can be roughly categorized into two groups, i.e., rule-based and data-driven methods.

\subsubsection{Rule-based Method}
The idea of the rule-based approach is to collect a set of gesture units and design specific rules that map a speech to a sequence of gesture units \cite{Kipp2004_Gesture, huang2012robot, softbank2018naoqi, cassell2004beat}. \citet{wagner2014gesture} have an excellent review of these methods. The results of the rule-based methods are generally highly explainable and controllable. However, the gesture units and rules typically have to be created manually, which can be costly and inefficient for complex systems.

\subsubsection{Data-driven Method}
Early research in data-driven method learns the rules embedded in data and combines them with predefined animation units to generate new gestures. For example, \citet{kopp2006towards, levine2010gesture} use probabilistic models to build correspondence between speech and gestures. \citet{Neff2008Gesture} build a statistical model to learn the personal style of each speaker. The model is combined with the input text tagged with the theme, utterance focus, and rheme to generate gesture scripts, which are then mapped to a sequence of gestures selected from an animation lexicon. \citet{chiu2015predicting} train a neural classification model to select a proper gesture unit based on the speech input. More recent research has started to take advantage of deep learning and trains  end-to-end models using raw gesture data directly, which frees the manual efforts of designing the gesture lexicon and mapping rules. Gestures can be synthesized using deterministic models such as multilayer perceptron (MLP) \cite{kucherenko2020gesticulator}, recurrent neural networks \cite{hasegawa2018evaluation, yoon2019robots, yoon2020speech, bhattacharya2021speech2affectivegestures, liu2022learning}, convolutional networks \cite{habibie2021learning}, and transformers \cite{9417647}, or by learning generative models such as normalizing flow \cite{alexanderson2020style}, VAEs \cite{li2021audio2gestures, xu2022freeform}, and learnable noise codes \cite{qian2021speech}. Our method is also a data-driven framework. We learn the motion generator and the mapping between the speech and gestures from data using a combined network structure of the vector quantized variational autoencoder (VQ-VAE)~\cite{oord2017neural} and LSTM. To capture the rhythmic and semantic correspondences between the speech and gestures, we propose a multi-stage architecture that explicitly models the rhythm and semantics in different stages. An earlier system proposed by \citet{kucherenko2021speech2properties2gestures} shares a similar high-level architectural design to our framework. However, there are two key differences: (a) our method is essentially an unsupervised learning approach, which learns the gesture lexeme, style code, and the generator directly from the data without detailed annotations; and (b) our system employs an explicit beat-based segmentation scheme which is shown to be effective in ensuring temporal coherence between the speech and the gesture.

\subsection{Multi-Modal Data Processing}
Co-speech gesture generation is a cross-modal process involving audio, text, motion, and other information related to the speaker and the content of the speech. The representation and alignment of each modality are essential for high-quality results~\cite{8269806}. Mel-spectrogram and MFCC acoustic features are commonly used as audio features \cite{alexanderson2020style, qian2021speech, kucherenko2020gesticulator}, typically resampled into the same framerate of the motion. For the text features, pre-trained language models like BERT \cite{devlin2019bert,kucherenko2020gesticulator} and FastText~\cite{bojanowski2017enriching,yoon2020speech} have been used to encode text transcripts into frame-wise latent codes, where paddings, fillers, or empty words are inserted into a sentence to make the world sequence the same length as the motion \cite{kucherenko2020gesticulator, yoon2020speech}. Speaker's style and emotions can also be encoded by learnable latent codes \cite{bhattacharya2021speech2affectivegestures,yoon2020speech} and are resampled or padded to match the length of the speech. In this work, we employ a pre-trained speech model to extract audio features and fine-tune it using a contrastive learning strategy. We also utilize a BERT-based model to vectorize the text. These multi-modal data are then aligned explicitly using the standard approaches discussed above. Notably, a concurrent study \cite{liu2022learning} also extracts audio features using contrastive learning. Their framework considers the learning of the audio features as a part of the training of the gesture generator. Instead, our framework trains the audio encoder in a separate pre-training stage using only the audio data.

\begin{figure*}[t]
    \centering
    \includegraphics[width=\textwidth]{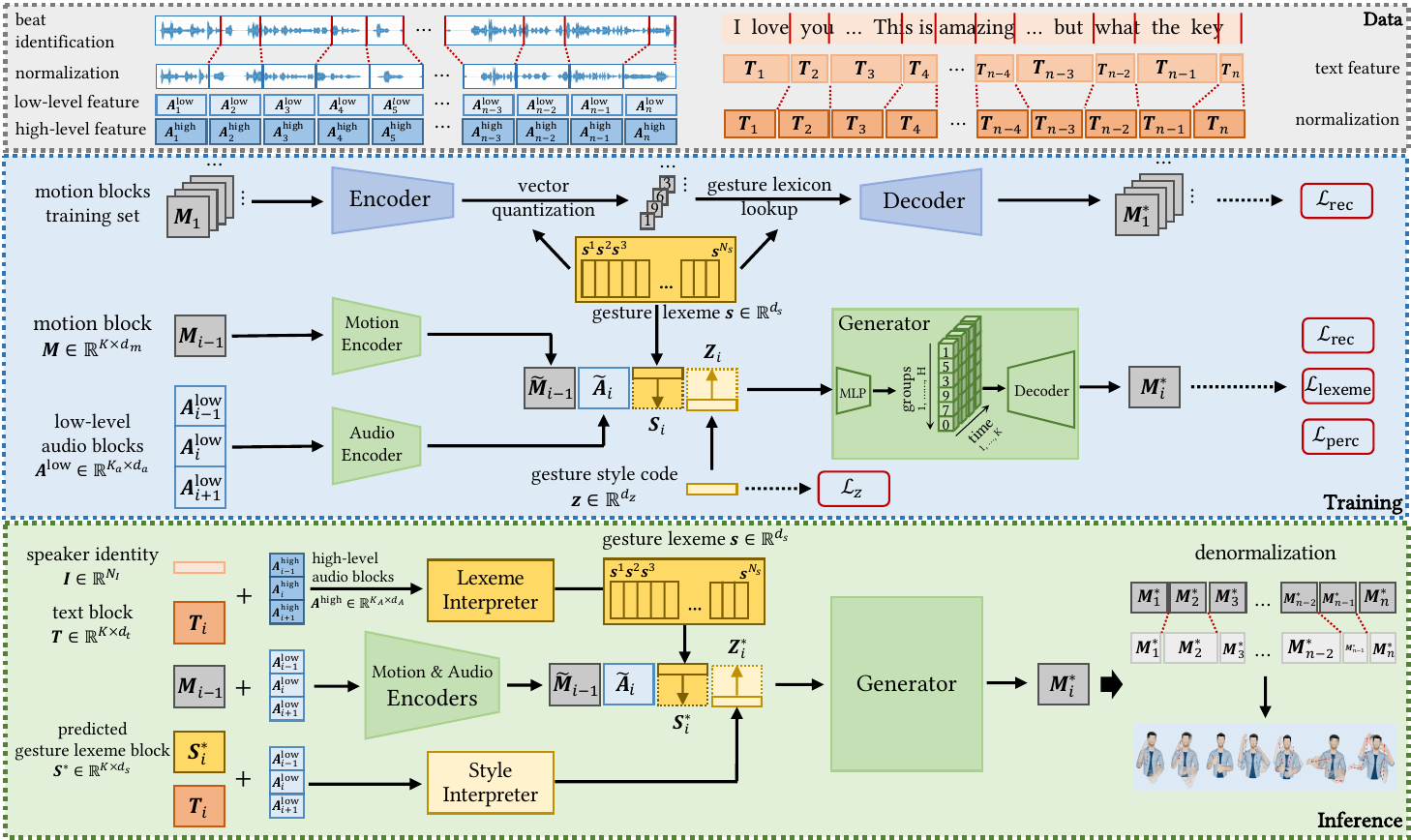}
    \caption{
    Our system is composed of three core components: 
    (a) the \emph{data} module preprocesses a speech, segments it into normalized blocks based on the beats, and extracts speech features from these blocks;
    (b) the \emph{training} module learns a gesture lexicon from the normalized motion blocks and trains the generator to synthesize gesture sequences, conditioned on the gesture lexemes, the style codes, as well as the features of previous motion blocks and adjacent speech blocks;
    and (c) the \emph{inference} module employs interpreters to transfer the speech features to gesture lexemes and style codes, which are then used by the learned generator to predict future gestures.
    }
    \Description{}
    \label{fig:system_overview}
\end{figure*}

\subsection{Evaluation of Motion Synthesis Models}
Evaluating the generated co-speech gestures is often difficult because the motion quality is a very subjective concept. Previous works have proposed several evaluation criteria. Wolfert et al.~\shortcite{wolfert2022evaluation} have made a comprehensive review of them. User studies are widely adopted to evaluate different aspects of motion quality, such as human-likeliness and speech-gesture matching \cite{alexanderson2020style, yoon2020speech, kucherenko2020gesticulator}, but can be expensive and hard to exclude uncontrolled factors. The absolute difference of joint positions or other motion features, such as velocity and acceleration between a reconstructed motion and the ground truth, is used by several works as an objective metric \cite{ginosar2019learning, joo2019towards, kucherenko2019analyzing}. However, this metric is not suitable for evaluating motions that are natural but not the same as the reference. Fr{\'e}chet Inception Distance (FID) \cite{heusel2017gans} is a widely used criterion in image generation tasks that measures the difference between the distributions of the dataset and generated samples in the latent space. It successfully reflects the perceptual quality of generated samples. Similarly, \citet{yoon2020speech} and \citet{qian2021speech} propose Fr{\'e}chet Gesture Distance (FGD) and Fr{\'e}chet Template Distance (FTD) metrics, respectively. These metrics measure the perceptual quality of generated gestures. In this paper, we compare our framework with several baseline methods using both user studies and objective metrics like FGD. We further propose a simple but effective rhythmic metric to measure the percentage of matched beats by dynamically adjusting the matching threshold, which provides a more informative picture of the rhythm performance.
\section{System Overview}
\label{sec:system_overview}
Our goal is to synthesize realistic co-speech upper-body gestures that match a given speech context both temporally and semantically. To achieve this goal, we build a system using neural networks that takes speech audio as input and generates gesture sequences accordingly. Additional speech modalities, such as text and speaker identity, will also be considered by the system when available to enhance semantic coherence and generate stylized gestures.

A gesture motion consists of a sequence of gesture units, which can be further broken down into a number of gesture phases that align with intonational units, such as pitch accents or stressed syllables \cite{kendon2004GestureBook,Loehr2012Temporal}. The action in each of these gesture phases is typically a specific movement such as lifting a hand, holding an arm at a position, or moving both arms down together, which is often referred to as a \emph{gesture lexeme} by linguists~\cite{Neff2008Gesture,Kipp2004_Gesture,Webb1996_Linguistic}. It is also revealed in the literature that there are only a limited number of lexemes used in everyday conversation. These lexemes form a \emph{gesture lexicon}. A typical speaker may only use a subset of this lexicon and apply slight variations to the motion. We assume such variations cannot be inferred directly from the speech but can be characterized by some latent variables, which we refer to as the \emph{gesture style code}s. Our system then generates gestures in a hierarchical order. It first determines the sequence of gesture lexemes and style codes and then generates gestural moves based on these motion-related features and other speech modalities.

Our system processes the input speech in a block-wise manner. Considering the temporal and structural synchrony between the gesture and the speech, we leverage a segmentation that aligns with the rhythm of the speech to ensure temporal coherence between the two modalities. Specifically, our system extracts beats from the input speech based on audio onsets and segments the speech into short clips at every beat. These clips are then time-scaled and converted into normalized blocks with the same length. We extract features at multiple levels for each block, where the high-level features are translated into a gesture lexeme, and the low-level features determine the style code. The generated gesture motions are then denormalized to match the length of the input speech.

As illustrated in \fig\ref{fig:system_overview}, our system consists of three core components: (a) the \emph{data} module preprocesses a speech, segments it into normalized blocks based on the beats, and extracts speech features from these blocks; (b) the \emph{training} module learns a gesture lexicon from the normalized motion blocks and trains the generator to synthesize gesture sequences, conditioned on the gesture lexemes, the style codes, as well as the features of previous motion blocks and adjacent speech blocks; and (c) the \emph{inference} module employs interpreters to transfer the speech features to gesture lexemes and style codes, which are then used by the learned generator to predict future gestures.

We train our system on a speech-gesture dataset with accompanying text and speaker identity (ID). The gesture lexicon is constructed using unsupervised learning based on a vector quantized variational autoencoder (VQ-VAE)~\cite{oord2017neural}. The generator is trained as an autoregressive encoder-decoder network, where we use an LSTM-based decoder combined with a vector quantized encoder to generate gesture motions. We train two separate interpreters to translate speech features into gesture lexemes and style codes, respectively. These interpreters can work with only the audio features and can be retrained to accept other speech modalities. In the following sections, we will provide details about these components and how they are trained in our system.

\section{Data Preparation}
\label{sec:data_preparation}
The \emph{data} module of our system preprocesses an input speech, segments it into uniform blocks based on speech rhythms, and extracts features that will be used to generate co-speech gestures. In this section, we first introduce the representations of different speech modalities and gesture motion and then describe details of the data preprocessing.

\subsection{Representation of Speech Modalities}

\subsubsection{Motion Representation}
\begin{figure}[t]
    \centering
    \includegraphics[width=0.9\linewidth]{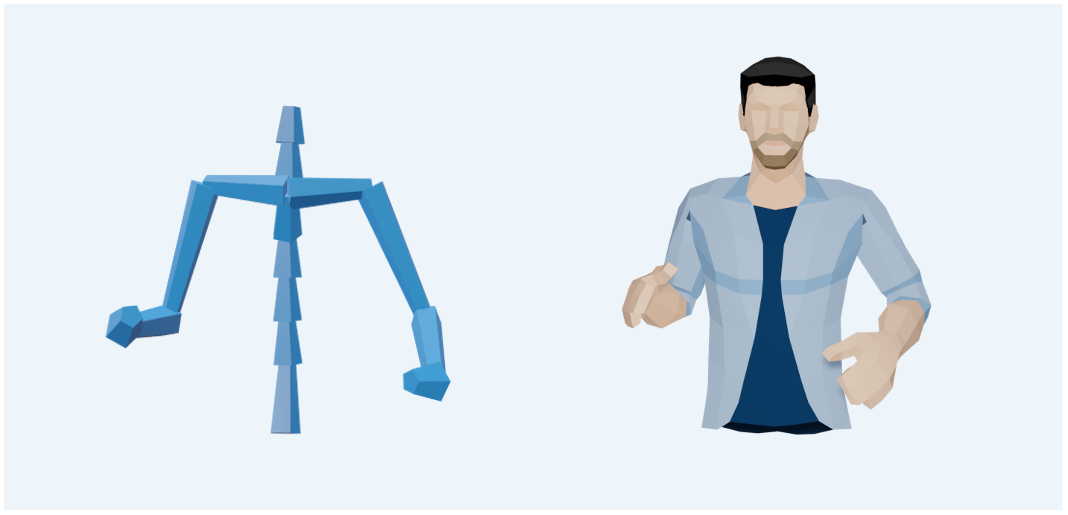}
    \caption{The character model used in our system.}
    \Description{}
    \label{fig:character_model}
\end{figure}
We focus on upper-body gestures in this work. Our system employs a character model consisting of $16$ upper-body joints, including a rotational root, as shown in \fig\ref{fig:character_model}. A gesture pose is then represented as a list of joint rotations, parameterized using the exponential map, in the hierarchical order. We use $\vect{m}_k\in\mathbb{R}^{d_m}$ to represent the gesture pose at frame $k$, and a clip of gestures is  represented collectively as $\vect{M}=\{\vect{m}_1,\dots,\vect{m}_K\}$, where $K$ is the number of frames. We retarget training motions onto this model by copying the rotations of corresponding joints. The translation and the rotation around the vertical axis are excluded from the root joint, ensuring a normalized body orientation.

\subsubsection{Text Representation}
Text transcription is an important speech modality that provides high-level linguistic information in a compact format. It is typically given as a word sequence, where the number of words per unit time can vary depending on the speed of the speech. Following \cite{kucherenko2020gesticulator}, we align the words to the speech and convert the text into frame-level features to overcome this issue, which is done using an off-the-shelf text-speech alignment tool combined with a pre-trained language model.

Text-speech alignment is a standard technique in the field of speech synthesis. In our system, we employ Montreal Forced Aligner (MFA) \cite{mcauliffe2017montreal} for this task, which pinpoints the beginning and end frames of every word in the speech. MFA also identifies silences and represents them as empty words. Since a speaker typically stops gesticulating in a long silence~\cite{graziano2018silence}, our system records those silences and uses them during training to reproduce such behaviors, as will be detailed later.

We then pass the text and the empty words into BERT~\cite{devlin2019bert}, a popular and powerful pre-trained language model, to extract a high-level representation of the text. BERT computes an encoding vector for each word in an input sentence, which is then repeated and used for all the frames that the word occupies. We represent these word vectors collectively as $\vect{T}=\{\vect{t}_1,\dots,\vect{t}_K\}$ for a speech clip of $K$ frames, where each $\vect{t}\in\mathbb{R}^{d_t}$.

\subsubsection{Audio Representation}
\label{subsubsec:audio_representation}
\begin{figure}[t]
    \centering
    \includegraphics[width=\linewidth]{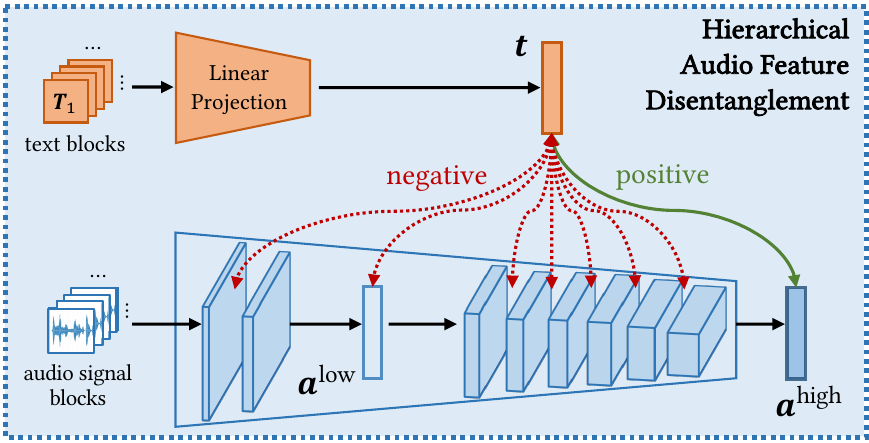}
    \caption{A contrastive learning task is performed to disentangle multi-level audio features. We use the text feature $\vect{t}$ as the anchor of this learning. The highest-level audio feature $\vect{a}^{\eqword{high}}$ is considered as the positive sample, while the features of the lower levels are all treated as negative samples. After training, both $\vect{a}^{\eqword{high}}$ and the feature extracted at the second level, $\vect{a}^{\eqword{low}}$, are used for gesture generation.}
    \Description{}
    \label{fig:audio_feature_extraction}
\end{figure}
Many recent studies use deep encoders to extract audio features from raw audio signals or audio spectrograms, where only the features extracted at the last layer of the encoder are used to generate gestures~\cite{yoon2020speech,alexanderson2020style,kucherenko2020gesticulator,qian2021speech,li2021audio2gestures}. Such a configuration potentially encourages the encoder to mix information from multiple levels into the same feature, which can be difficult to disentangle in the downstream generation tasks.

In our system, we propose to decouple the multi-level audio features in the encoder and use them in different scenarios. We assume the high-level features correspond to the speech semantics that determines the gesture lexemes, while the low-level features relate to the other audio information and can be used to control the {gesture styles}. As shown in \fig\ref{fig:audio_feature_extraction}, we employ a pre-trained speech model, {vq-wav2vec}~\cite{baevski2020vq-wav2vec}, to extract audio features from raw audio signals and fine-tune it using a contrastive learning strategy.

The encoder of vq-wav2vec has $L=8$ convolutional layers. When taking a block of audio signals of $K$ frames, $\vect{A}$, as input, this encoder produces a representation $\vect{a}_k$ for each frame $k$ of the audio. In this computation, the outputs of every layer can be considered as a set of multi-level features $\{\vect{a}_k^{l}\}, {l=1,\dots,L}$, and notably, $\vect{a}_k^{L}=\vect{a}_k$. We then encourage the highest-level feature $\vect{a}_k^{L}$ to match the speech content and push apart the features of the lower levels $\{\vect{a}_k^{l}\}, l<L$ to capture crucial content-irrelevant information. Specifically, we utilize the contrastive loss
\begin{equation}
    \mathcal{L}_{\eqword{cont}} = -\log{\frac{\exp{(\simloss(\tildevect{t}_{k}, \tildevect{a}^{L}_{k})/\tau)}}{\sum^{K}_{i=1}\sum_{l=1}^{L}\exp{(\simloss(\tildevect{t}_k, \tildevect{a}^{l}_{i})}/\tau)}},
\end{equation}
where the text feature $\vect{t}_k$ is extracted from the speech transcription, the $\simloss(\cdot,\cdot)$ function computes the cosine similarity between two vectors as
\begin{equation}
    \simloss(\tildevect{t},\tildevect{a})=\frac{\tildevect{t} \cdot \tildevect{a}}{\norm{\tildevect{t}}\norm{\tildevect{a}}},
\end{equation}
and $\tau$ is the temperature hyperparameter. All the feature vectors are projected into the same vector space using learnable linear projections $\tildevect{t}=F_t(\vect{t})$ and $\tildevect{a}_k^l=F_a^l(\vect{a}_k^l)$, $l=1,\dots,L$, respectively. Notably, we consider the highest-level audio feature of the current frame as the positive example and audio features of the other levels and the other frames as the negative examples in this contrastive learning process.

This contrastive learning strategy is partially inspired by the HA2G model proposed by Liu et al.~\shortcite{liu2022learning}. However, unlike their approach, which considers contrastive learning as a part of the training of the gesture generator, we train the audio encoder in a separate pre-training stage using only the speech data. After the training, the features extracted at the second and the last layers of the encoder, represented by $\vect{a}^{\eqword{low}}\in\mathbb{R}^{d_a}$ and $\vect{a}^{\eqword{high}}\in\mathbb{R}^{d_A}$, respectively, are then used in different training and inference stages in the downstream generation task. They can be represented collectively as $\vect{A}^{\eqword{low}}$ and $\vect{A}^{\eqword{high}}$ for a speech clip. Although the gesture motions are not considered here, we find that the results of this encoder still demonstrate correlations between the high-level audio features and the gestures. We will discuss these results later in Section~\ref{subsec:ablation_study}.

\subsubsection{Identity Representation}
Similar to previous studies \cite{yoon2020speech,bhattacharya2021speech2affectivegestures}, our system can leverage the speaker identity (ID) to help distinguish different gesture styles and achieve stylized gesture generation. We represent each speaker as a one-hot vector $\vect{I}\in\{0,1\}^{N_I}$, where $N_I$ is the number of speakers in a dataset.
\begin{figure}[t]
    \centering    
    \begin{subfigure}[t]{0.47\linewidth}
        \centering
        \caption*{Trinity Gesture Dataset}
        \includegraphics[width=\linewidth]{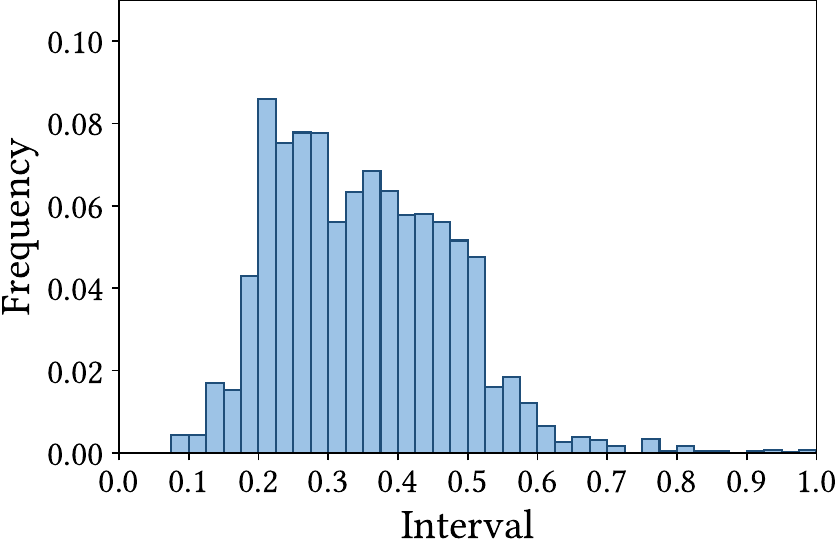}
   \end{subfigure}
   \hspace{\fill}
    \begin{subfigure}[t]{0.47\linewidth}
        \centering
        \caption*{TED Gesture Dataset}
        \includegraphics[width=\linewidth]{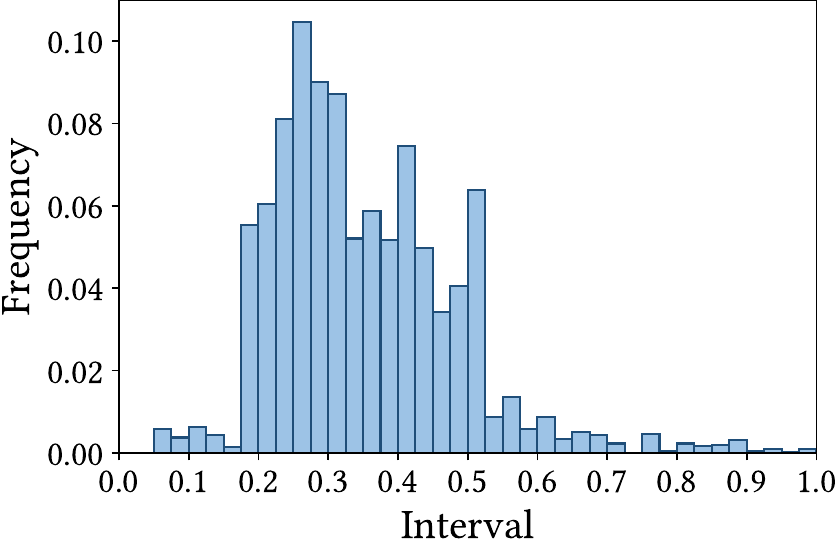}
   \end{subfigure}
    \caption{Distribution of time intervals between consecutive audio onsets in two open-source speech-gesture datasets.}
    \Description{}
    \label{fig:onset_intervals}
\end{figure}
\subsection{Rhythm-Based Speech Segmentation}
\label{subsec:beat_alignment}
In this section, we describe how our system segments and normalizes an input speech into uniform blocks. This procedure is crucial for generating a gesture motion that is temporally synchronized with the rhythm of the speech. To that end, our system first identifies beats in the input audio, which generally corresponds to phonetic properties such as stress or accent, then segments the speech at every beat and time-scales the audio to the same length.

\subsubsection{Beat Identification}
\label{subsubsec:onset_identification}
Rhythm can be characterized by a pattern of beats. In music-related tasks, such as dance generation~\cite{chen2021choreomaster,aristidou2021rhythm}, identifying beats using the onsets of audio signals is a standard technique~\cite{bello2005onsetTutorial,ellis2007beat}, where off-the-shelf tools such as librosa library~\cite{mcfee2015librosa} can be employed to extract those audio features.

However, unlike the rhythm in music that is typically consistent over time, the pattern of beats in a speech can vary significantly according to the context and pace of the speech. Taking a close look at the time intervals between consecutive audio onsets in our training dataset, we notice that the majority of those intervals fall within a range roughly between $D_m=0.2\sim{}0.3$ seconds and $D_M=0.5\sim{}0.6$ seconds, as illustrated in \fig\ref{fig:onset_intervals}, though the actual values of $D_m$ and $D_M$ may vary among datasets depending on the personality of the speakers and the language they speak. We also observe that the time intervals shorter than $D_m$ are often caused by noise, filler words, or stuttering. On the other hand, the intervals that are excessively long often correspond to pauses or silent periods.

Based on these observations, our system employs a simple heuristic strategy to identify beats based on the audio onsets. An onset will be recognized as a beat unless the time interval between it and the previous beat is shorter than $D_m$, in which case the onset will be ignored. If an interval is longer than $D_M$, we will insert a necessary number of \emph{pseudo-beat}s to make the duration of every new interval within the range $[D_m, D_M]$. More specifically, we insert a pseudo-beat at the first frame which is $D_m$ seconds away from any preceding beat and where the volume of the audio is greater than a threshold $\tilde{I}_a$. Other pseudo-beats are then added recursively in the same way. We set the threshold $\tilde{I}_a$ as the average volume of the environmental noise. If the entire interval is quieter than $\tilde{I}_a$, a minimal number of pseudo-beats will be placed evenly in it so that each new interval is shorter than $D_M$.

\subsubsection{Normalization}
\label{subsubsec:normalization}
Our system then segments the speech into short clips at every beat. These clips are then time-scaled into uniform blocks of length $D_M$. The speech modalities are segmented and time-scaled as well in this process. For the motion, $\vect{M}$, and text representation, $\vect{T}$, of a clip, we resample the corresponding features to match the new length. The audio is processed with additional care, where we use the TSM (Time-Scale Modification) algorithm to change the duration of the audio while preserving the pitch. The audio features $\vect{A}^{\eqword{low}}$ and $\vect{A}^{\eqword{high}}$ are then recomputed for the time-scaled audio blocks. The speaker ID $\vect{I}$ is a constant of the whole speech, which will not be changed during the normalization.
\section{Gesture Generation}
\label{sec:gesture_generation}
The generator module is the core component of our system. It synthesizes realistic gesture motions according to a sequence of gesture lexemes, the corresponding style codes, and the low-level features of the audio. In this section, we first introduce how we construct the gesture lexicon and then describe the design and training of the gesture generator.

\subsection{Construction of Gesture Lexicon}
\label{subsec:gesture_style_embedding}
As revealed in several pieces of literature in linguistics \cite{Neff2008Gesture,Kipp2004_Gesture,Webb1996_Linguistic}, only a limited number of lexemes are used in everyday conversation. We assume that each lexeme corresponds to a specific motion category. Our goal is then to extract those motion categories from a large gesture dataset. To achieve this goal, we employ the vector quantized variational autoencoder (VQ-VAE) model \cite{oord2017neural} to learn a categorical representation of the motion and construct the gesture lexicon.

VQ-VAE has been widely used to learn categorical spaces in many successful temporal models \cite{prafulla2020jukebox, baevski2020vq-wav2vec, yan2021videogpt,ramesh2021DALLE}. Similar to a regular autoencoder, a VQ-VAE also has an encoder-decoder structure but quantizes the latent space using a discrete codebook. The codebook consists of a list of vectors and their associated indices. The output of the encoder network is compared to every vector in the codebook, where the vector that is the closest in Euclidean distance is considered to be the latent representation of the input and will be fed to the decoder. The training of a VQ-VAE is achieved by pulling together the latent code of input and its corresponding codebook vector.

As illustrated in \fig\ref{fig:system_overview}, we construct the gesture lexicon by learning the categorical representations of the normalized motion blocks using VQ-VAE. Following \cite{oord2017neural}, the loss function is defined as
\begin{align}
    \mathcal{L}_{\eqword{lexicon}} &= \lVert \vect{M}-\mathcal{D}({\vect{s}}) \rVert_2^2 \nonumber\\
    &+  w_{\alpha}\lVert \mathcal{E}(\vect{M}) -\stopgrad({\vect{s}}) \rVert_2^2 
    +  w_{\beta}\lVert \stopgrad(\mathcal{E}(\vect{M})) - {\vect{s}} \rVert_2^2,
    \label{eqn:vq_vae_loss}
\end{align}
where
\begin{equation}
    {\vect{s}} = \argmin_{\vect{s}'\in \mathcal{S}} \lVert \vect{s}' - \mathcal{E}(\vect{M}) \rVert_2, \label{eqn:codebook_lookup}
\end{equation}
$\vect{M}$ is a normalized motion block, $\mathcal{E}$ and $\mathcal{D}$ represent the encoder and decoder, respectively, $\stopgrad$ stands for the \emph{stop gradient} operator that prevents the gradient from backpropagating through it, $\mathcal{S}$ represents the codebook, or the lexicon, and $\vect{s}$ is a codebook vector, or a lexeme. The first term of \eqn\eqref{eqn:vq_vae_loss} penalizes the reconstruction error, while the other two terms pull together the latent representation of motion $\vect{M}$ and its corresponding lexeme. Notably, since $\mathcal{S}$ is discrete, the $\argmin$ operator in \eqn\eqref{eqn:codebook_lookup} does not generate a gradient. The gradient of the reconstruction error with respect to the latent code is passed unaltered to the encoder during the backward pass as suggested in~\cite{oord2017neural}.
\begin{figure}[t]
    \centering
    \begin{subfigure}[t]{0.47\linewidth}
        \centering
        \includegraphics[width=\linewidth]{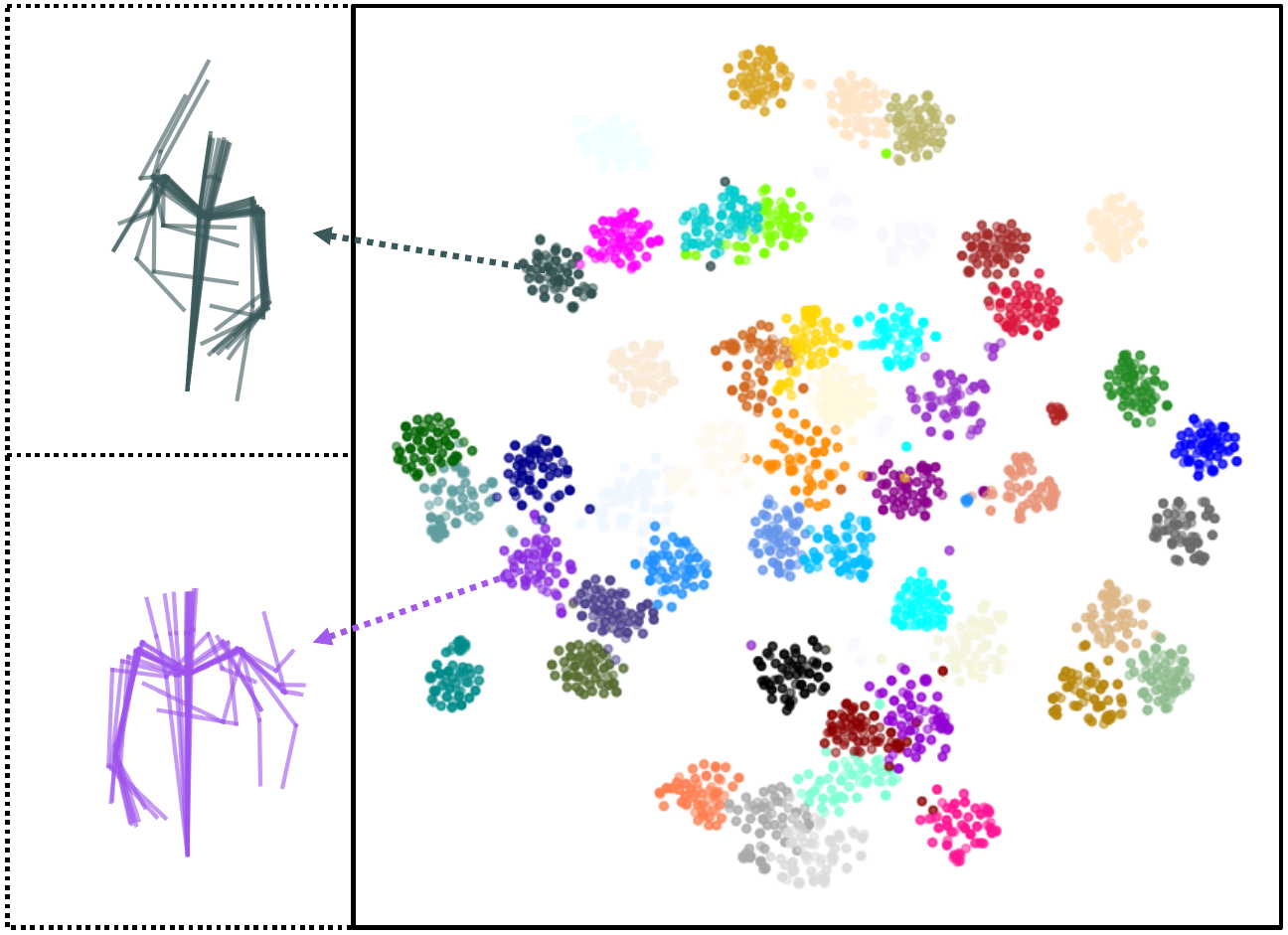}
        \caption{Trinity Gesture dataset.}
        % \label{fig:fig3a}
    \end{subfigure}
    \hspace{\fill}
    \begin{subfigure}[t]{0.47\linewidth}
        \centering
        \includegraphics[width=\linewidth]{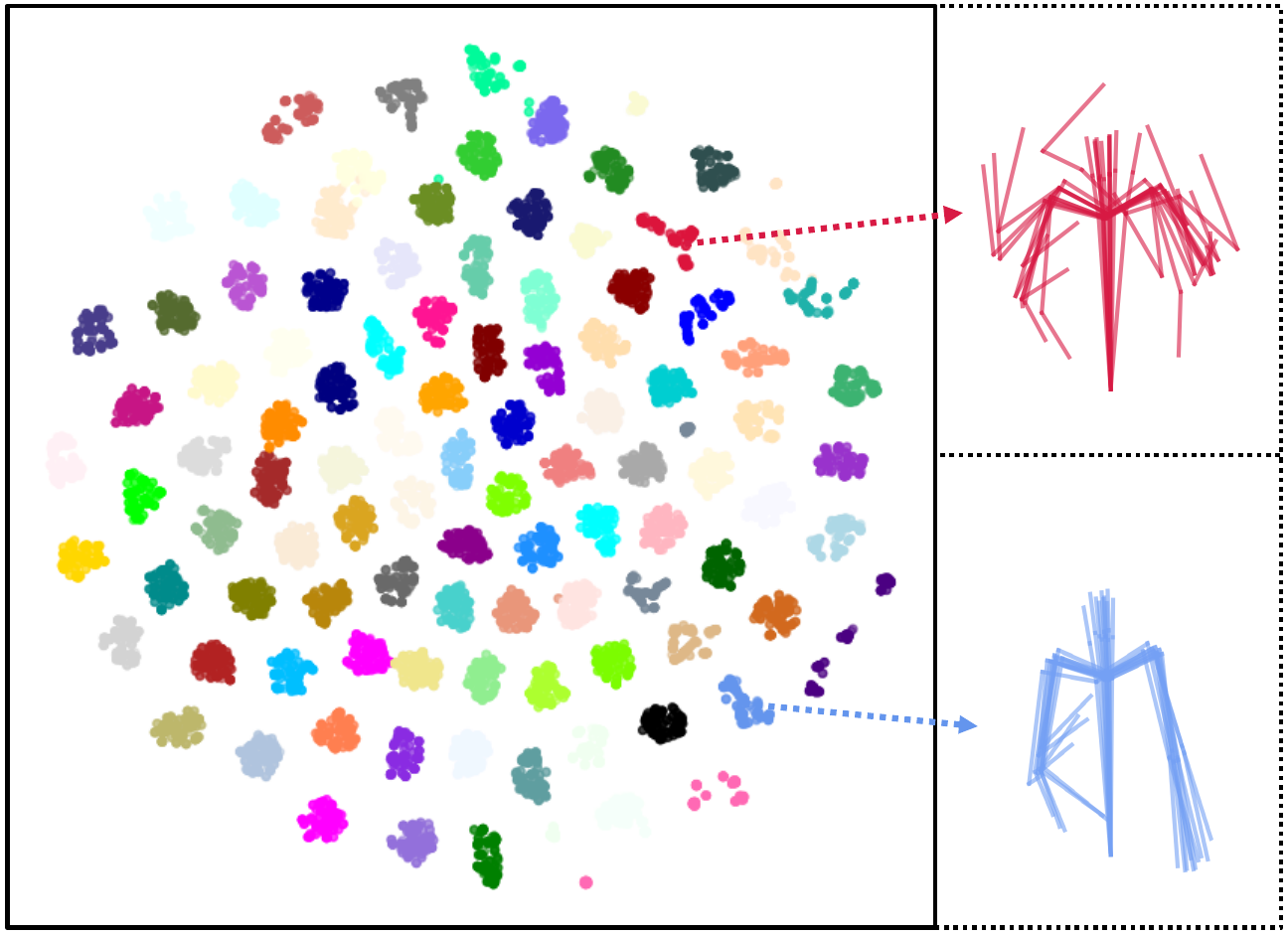}
        \caption{TED Gesture dataset.}
        % \label{fig:fig3b}
    \end{subfigure}
    \caption{t-SNE visualization of gesture lexicons. Each color stands for a gesture lexeme. (a) lexicon learned on the Trinity Gesture dataset with 50 lexemes. (b) lexicon learned on the TED Gesture dataset with 100 lexemes.}
    \Description{}
    \label{fig:gesture_lexicon}
\end{figure}

We train the VQ-VAE on each dataset in a separate pre-training stage. The encoder is a multi-layer network consisting of four 1-D convolutional layers followed by a fully connected layer, which encodes a motion block into a vector $\vect{s}\in\mathbb{R}^{d_s}$. The decoder is a mirror of the encoder structurally. The size of the lexicon is a hyperparameter, which is chosen empirically based on the size and complexity of the dataset. \fig\ref{fig:gesture_lexicon} shows the t-SNE visualization of the training results on two speech-gesture datasets, along with sample gestures of several lexemes. Once learned, the gesture lexicon and the gesture lexeme of every motion block are fixed and used by the generator and interpreters in both the training and inference stages.

\subsection{Architecture of Generator}
As illustrated in \fig\ref{fig:system_overview}, the generator module of our system is an autoregressive encoder-decoder network, where a new motion block is conditioned on not only the input speech but also the preceding block, the gesture lexeme, and the style code. More specifically, the generation of a motion block $\vect{M}_{i}$ can be formulated as
\begin{equation}
    \vect{M}_{i}^*=\mathcal{G}(\tildevect{M}_{i-1},\tildevect{A}_{i},\vect{s}_i,\vect{z}_i,\vect{P}) \label{eqn:generator_block},
\end{equation}
where $\tildevect{M}_{i-1}$ represents the features extracted from the preceding motion block, $\tildevect{A}_{i}$ stands for the representation of the input audio, $\vect{s}_i$ and $\vect{z}_i$ are the gesture lexeme and style code of the new block, respectively. Note that we use an asterisk~($*$) to indicate a generated quantity. All these motion and feature blocks have the same length of $K$ frames, where $\vect{s}_i$ and $\vect{z}_i$ are repeated and stacked into the corresponding blocks, represented as $\vect{S}_i$ and $\vect{Z}_i$, respectively.

We extract the motion feature $\tildevect{M}_{i-1}\in\mathbb{R}^{K\times{}d_{\tilde{m}}}$ as 
\begin{equation}
    \tildevect{M}_{i-1} = \mathcal{E}_M(\vect{M}_{i-1}),
\end{equation}
where the encoder $\mathcal{E}_M$ is a 1-D convolutional network with three layers.
The audio feature $\tildevect{A}_{i}\in\mathbb{R}^{K\times{}d_{\tilde{a}}}$ is computed using three consecutive audio blocks
\begin{equation}
    \tildevect{A}_{i} = \mathcal{E}_A(\vect{A}^{\eqword{low}}_{i-1},\vect{A}^{\eqword{low}}_{i},\vect{A}^{\eqword{low}}_{i+1})
\end{equation}
to allow the generator to prepare for the future gestures. 
Notably, the original duration of an audio block is characterized by the onset interval $[D_m, D_M]$, which is typically $[0.2s, 0.5s]$ in our experiments. Thus the temporal window of this encoder is roughly $[0.6s, 1.5s]$.
Each $\vect{A}^{\eqword{low}}$ is the low-level feature pre-computed from the raw speech audio. We assume that $\vect{A}^{\eqword{low}}$ already captures necessary information and use a simple network consisting of one fully connected layer as the encoder $\mathcal{E}_A$.

In the training stage, the gesture lexeme $\vect{s}_i$ of each motion block is determined during the construction of the gesture lexicon, while the style code $\vect{z}_i$ is a learnable variable that will be trained along with the generator. In the inference stage, both $\vect{s}_i$ and $\vect{z}_i$ are provided by the interpreters, as will be discussed below.

In addition to these features, we include a positional encoding block, $\vect{P}\in\mathbb{R}^{K\times{}d_{P}}$, to let the generator know the progress of the generation in a motion block, which is a standard component of transformers~\cite{Vaswani2017_Attentiona} and many sequence generation tasks~\cite{Harvey2020_motionInBetween}. For our normalized blocks with $K$ frames, we compute $\vect{P}=\{\vect{p}_1,\dots,\vect{p}_K\}$ as
\begin{align}
    \vect{p}_{2k} = \sin\left(\frac{K}{\beta^{2k/d_{P}}}\right) \quad
    \vect{p}_{2k+1} = \cos\left(\frac{K}{\beta^{2k/d_{P}}}\right),
    \label{eqn:pos_enc}
\end{align}
where $d_{P}$ is the dimension of the encoding and $\beta=10,000$ is a constant controlling the rate of change in frequencies along the embedding dimensions.

The generator $\mathcal{G}$ consists of an MLP-based encoder followed by an LSTM-based decoding network. Inspired by the successful systems in generating sequential output \cite{richard2021meshtalk,oord2017neural}, we quantized the latent space of the encoder into $H$ groups of $C$-way categories, which provides $C^H$ different configurations. We use $H=64$ and $C=128$ to ensure a large enough categorical space. In addition, we leverage Gumbel-softmax~\cite{jang2017categorical} to convert a latent code into a codebook vector, which can be viewed as a differentiable sampling operator for the discrete codebook search.

\subsection{Training}
We train the generator $\mathcal{G}$, the encoders $\mathcal{E}_M$ and $\mathcal{E}_A$, and the learnable style codes $\{\vect{z}_i\}$ by minimizing a combination of loss terms:
\begin{equation}
    \mathcal{L}_{\eqword{gen}} 
    = w_{\eqword{rec}}\mathcal{L}_{\eqword{rec}}
    + w_{\eqword{perc}}\mathcal{L}_{\eqword{perc}}
    + w_{\eqword{lexeme}}\mathcal{L}_{\eqword{lexeme}}
    + w_{{z}}\mathcal{L}_{{z}}.\label{eqn:generator_loss}
\end{equation}
The reconstruction loss 
\begin{equation}
    \mathcal{L}_{\eqword{rec}}=\lVert \vect{M}_i - \vect{M}_i^* \rVert_2^2   
\end{equation}
is simply the MSE loss between the generated motion block and the ground truth. We additionally include a perceptual loss to ensure the similarity between the generated motion and the ground truth in the feature level as well, which is defined as
\begin{equation}
    \mathcal{L}_{\eqword{perc}} = \lVert \mathcal{E}(\vect{M}_i) - \mathcal{E}(\vect{M}_i^*) \rVert_2^2 ,
\end{equation}
where $\mathcal{E}$ is the motion encoder pre-trained in Section~\ref{subsec:gesture_style_embedding}.

We assume that the gesture lexeme determines the type of gesture motion, and the other speech modalities only affect the motion variations. To enforce this assumption, we develop a new perceptual loss, namely the lexeme loss. We first generate a number of new motion blocks using the current gesture lexeme but random sets of other features, 
\begin{equation}
    \vect{M}^{*}_j = \mathcal{G}(\tildevect{M}_{j-1}, \tildevect{A}_{j}, \vect{s}_i, \vect{z}_j, \vect{P}) ,
\end{equation}
where $\tildevect{M}_{j-1}$, $\tildevect{A}_{j}$, and $\vect{z}_j$ correspond to a random speech block $j$. Then the lexeme loss is defined as 
\begin{equation}
    \mathcal{L}_{\eqword{lexeme}}=\frac{1}{N_J}\sum_{j\in{}J}\lVert \vect{s}_i - \mathcal{E}(\vect{M}_j^*) \rVert_2^2 ,
\end{equation}
where $J$ is a random subset of all the motion blocks in the training dataset. The size of $J$, $N_J$, is chosen based on the size of the dataset.

Lastly, we regularize the learning of the style code by applying a KL-divergence loss
\begin{equation}
    \mathcal{L}_{z} = D_{KL}(\mathcal{N}(\mu_z, \sigma^2_z) \Vert \mathcal{N}(0, 1)),
\end{equation}
where $\mu_z$ and $\sigma^2_z$ are the mean and variance vectors of the style codes in a mini-batch, respectively.
\section{Co-speech Gesture Inference}
\label{sec:co-speech_gesture_inference}
When given a speech as input, our system segments it into normalized feature blocks $\{\vect{A}_i^{\eqword{low}}, \vect{A}_i^{\eqword{high}}, \vect{T}_i, \vect{I}\}$ and then generates motion blocks $\{\vect{M}_i^*\}$ recursively, where 
\begin{align}
    \vect{M}_i^* = \mathcal{G}\left(
        \mathcal{E}_M\left(\vect{M}_{i-1}^*\right), 
        \mathcal{E}_A\left(\vect{A}^{\eqword{low}}_{i-1}, \vect{A}^{\eqword{low}}_{i}, \vect{A}^{\eqword{low}}_{i+1}\right), 
        \vect{s}_i^*, \vect{z}_i^*, \vect{P}
        \right),
    \label{eqn:gesture_inference}
\end{align}
and $\mathcal{G}, \mathcal{E}_M, \mathcal{E}_A$ are the components of the learned gesture generator. The generated motion blocks are then denormalized to their original length in the input speech, producing a realistic co-speech gesture animation. Note that we again use the asterisk~($*$) to indicate a computed quantity that is not provided directly in the speech. 

All the variables in \eqn\eqref{eqn:gesture_inference} are known except the gesture lexeme $\vect{s}^*$ and style code $\vect{z}^*$. As shown in \fig\ref{fig:system_overview}, our system learns two interpreters to compute them: the \emph{lexeme interpreter} $\mathcal{P}_s$ translates high-level speech features into the gesture lexemes $\vect{s}^*$, and the \emph{style interpreter} $\mathcal{P}_z$ predicts the style code $\vect{z}^*$ according to the low-level speech features. 

\begin{figure}[t]
    \centering
    \includegraphics[width=\linewidth]{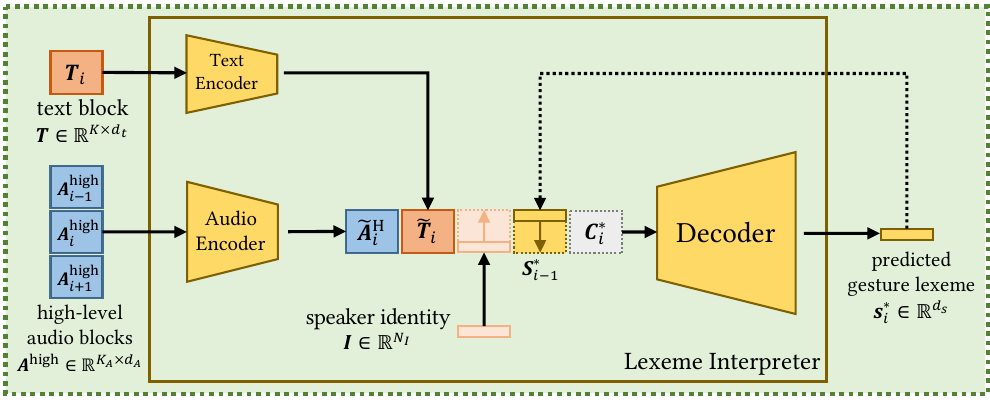}
    \caption{Architecture of the lexeme interpreter.}
    \Description{}
    \label{fig:interpreters}
\end{figure}

\subsection{Lexeme Interpreter}
\label{subsec:lexeme_interpreter}
As illustrated in \fig\ref{fig:interpreters}, the lexeme interpreter is formulated as
\begin{equation}
    \vect{s}_i^*=\mathcal{P}_s(\vect{s}_{i-1}^*, \tildevect{A}_i^{\eqword{H}}, \tildevect{T}_i, \vect{I}), 
    \label{eqn:lex_interpreter}
\end{equation}
which is conditioned on the gesture lexeme of the last motion block $\vect{s}_{i-1}^*$, and the high-level features of the current speech block. Like the generator, the high-level audio features $\tildevect{A}_i^{\eqword{H}}\in\mathbb{R}^{K\times{}d_{\tilde{A}}}$ are computed using three consecutive audio blocks
\begin{equation}
    \tildevect{A}_i^{\eqword{H}}=\mathcal{E}_A^{\eqword{lex}}(\vect{A}^{\eqword{high}}_{i-1},\vect{A}^{\eqword{high}}_{i},\vect{A}^{\eqword{high}}_{i+1}),
\end{equation}
where each $\vect{A}^{\eqword{high}}$ contains the high-level representation of the input speech audio, and the encoder $\mathcal{E}_A^{\eqword{H}}$ is a single-layer fully connected network. 
The text feature $\tildevect{T}_i\in\mathbb{R}^{K\times{}d_{\tilde{t}}}$ is also extracted from the text representation of the speech block as 
\begin{equation}
    \tildevect{T}_i=\mathcal{E}_T^{\eqword{lex}}(\vect{T}_i),
\end{equation}
where $\mathcal{E}_T^{\eqword{lex}}$ is again a single-layer network. % that changes the dimensionality of the text representation $\vect{T}_i$. 
Lastly, the one-hot representation of the speaker ID, $\vect{I}$, is repeated $K$ times and converted into a feature block.

Those feature blocks can then be concatenated together and fed to an LSTM-based decoder to predict the next gesture lexeme $\vect{s}_i^*$. However, considering that the lexemes are selected from the discrete gesture lexicon, we can convert this regression problem into a classification problem. Specifically, instead of directly evaluating \eqn\eqref{eqn:lex_interpreter}, we can let $\mathcal{P}_s$ predict the probability that $\vect{s}_i^*$ is a specific lexeme in the gesture lexicon, then the lexeme with the maximum likelihood will be considered as the result.

\subsection{Style Interpreter}
The style interpreter shares a similar structure with the lexeme interpreter. It computes $\vect{z}_{i}^*$ as
\begin{equation}
    {\vect{z}}_{i}^{*} = \mathcal{P}_z\left(
        \vect{z}_{i-1}^*,
        \vect{s}_{i}^*,
        \mathcal{E}_T^{\eqword{style}}\left(\vect{T}_i\right), 
        \mathcal{E}_A^{\eqword{style}}\left(\vect{A}^{\eqword{low}}_{i-1},\vect{A}^{\eqword{low}}_{i},\vect{A}^{\eqword{low}}_{i+1}\right)
        \right),
    \label{eqn:style_interpreter}
\end{equation}
which is conditioned on the last style code and the new gesture lexeme computed by the lexeme interpreter. The low-level audio representation $\vect{A}^{\eqword{low}}$ is used in the style interpreter.

\subsection{Audio-Only Inference}
\label{subsec:inference_based_on_audio_only}
Both the two interpreters can be reformulated to take only the speech audio as input, where the features related to the text representation $\vect{T}$, and optionally the speaker ID $\vect{I}$, will be removed from \eqn\eqref{eqn:lex_interpreter} and \eqref{eqn:style_interpreter}.

In practice, these audio-only interpreters allow cross-language gesture generation, where the speech audio in another language can be taken as input to synthesize realistic gestures without further training. For example, we can utilize a pre-trained model on an English dataset to generate gestures that accompany a Chinese speech. We will show related experiments in Section \ref{subsec:evaluation}.

\subsection{Training}
During the training of the generator, we have computed the gesture lexeme $\vect{s}_i$ and the style code $\vect{z}_i$ of every motion block in the training dataset. We then train the two interpreters using these results as the ground truth. We minimize the standard categorical cross-entropy loss to train the lexeme interpreter, while the MSE loss is used for the style interpreter.

\subsubsection{Silent Period Hint}
A speaker typically stops gesticulating during a silent pause \cite{graziano2018silence}. Such behaviors are often crucial to the naturalness of a co-speech gesture animation. However, we find that it is often difficult for a gesture generator to deal with silent periods well, even in recent successful systems such as \cite{alexanderson2020style,kucherenko2020gesticulator}. The speech-gesture datasets may lack necessary motion, and some specific generator models, such as LSTM, may exhibit generative inertia that makes it difficult to become stationary in time.

To solve this problem, we develop a new approach, which we refer to as the \emph{silent period hint}, to encourage the lexeme interpreter to compute a specific \emph{silent lexeme} that corresponds to a silent gesture when encountering a silent period. 
We check all the lexemes in the lexicon and label a number of stationary ones as the silent lexemes. Notably, the silent lexemes can be automatically labeled by finding such a lexeme corresponding to an empty text word. Then, when a training audio block is in a silent period, which can be detected by the data module of our system, we will force the lexeme interpreter to output the silent lexeme that is the nearest to the current lexeme in the latent space.
Moreover, a silent data augmentation is applied when training the generator. We find data blocks that contain empty words and randomly insert $0\sim{}10$ consecutive silent blocks after them. The silent block above includes four different features: (a) the audio feature is the environmental noise; (b) the style code is set to zero; (c) the gesture lexeme is the silent lexeme that is the nearest to the previous lexeme in the latent space; and (d) the motion is a stationary pose that is the same as the last frame of the previous motion block. In total, the amount of the inserted silent blocks accounts for $5\%$ of the whole training set.
\begin{figure}[t]
    \centering   
    \begin{subfigure}[t]{0.433\linewidth}
         \centering
         \includegraphics[width=\linewidth]{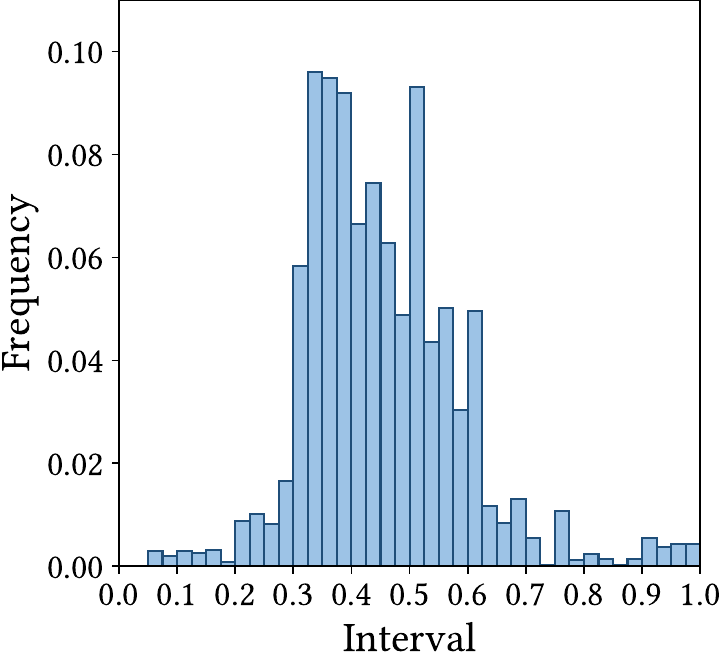}
         \caption{}
         \label{fig:onset_intervals_Chinese}
    \end{subfigure}
    \hspace{0.06\linewidth}
    \begin{subfigure}[t]{0.40\linewidth}
         \centering
         \includegraphics[width=\linewidth]{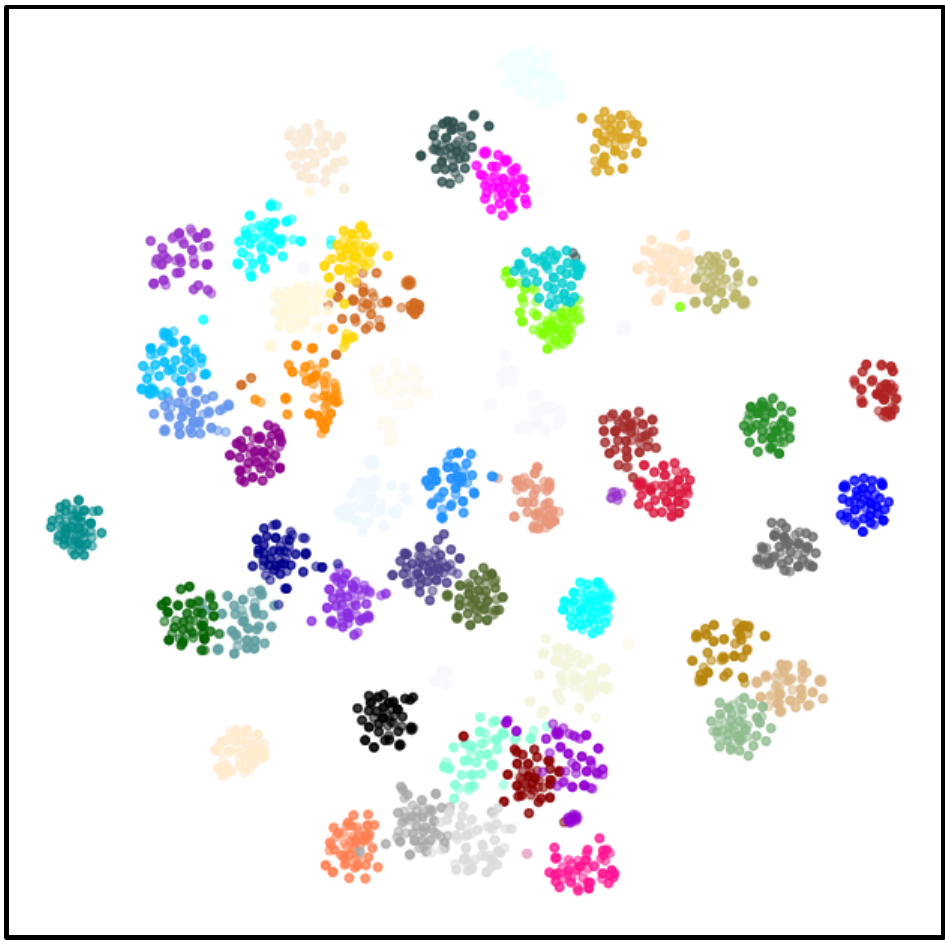}
         \caption{}
         \label{fig:lexicon_tsne_Chinese}
    \end{subfigure}
    \caption{Properties of our Chinese Gesture dataset. (a) Distribution of onset intervals; (b) t-SNE visualization of gesture lexicon.} 
    \Description{}
\end{figure}

\section{Results}
\label{sec:results}
In this section, we first introduce the setup of our system and then evaluate its performance, followed by quantitative and qualitative comparisons with other recent systems. Lastly, we do the ablation study to make an overall analysis of our system.

\subsection{System Setup}
\label{subsec:system_setup}
\subsubsection{Datasets}
\label{subsubsection:datasets}
Three speech-gesture datasets are used in this paper: the Trinity dataset \cite{ferstl2018investigating}, the TED dataset \cite{yoon2019robots}, and a Chinese dataset collected for this work.

Trinity Gesture dataset is a large database of speech and gestures jointly collected by \citet{ferstl2018investigating}. This dataset consists of $242$ minutes of motion capture and audio of one male actor talking on different topics. The actor’s motion was captured with a $20$-camera Vicon system and solved onto a skeleton with $69$ joints. In this paper, we use the official release version of \emph{The GENEA Challenge $2020$ Dataset} \cite{kucherenko2021large}, where $221$ minutes are used as training data, and the remaining $21$ minutes are kept for testing.

TED Gesture dataset \cite{yoon2019robots} is a 3D upper-body gesture dataset collected using 3D pose estimation from English TED videos. This dataset includes 3D upper-body gestures of the speakers, the aligned speech audios, and text transcripts. In total, there are $253,186$ data samples, $80\%$ of which are training samples, $10\%$ belong to the validation set, and the remaining $10\%$ are test samples. The duration of each data sample is $34$ frames at a rate of $15$ fps, so the total length of this dataset is about $97$ hours. Notably, we adapt our model to take 3D joint positions instead of rotations for the TED dataset. The generated gestures are also represented in 3D joint positions, which are then converted into joint rotations for visualization.

Additionally, we collected a $4$-hours ($80\%$ are the training data and $20\%$ are used for testing) Chinese Gesture dataset using the Noitom Perception Neuron Pro system. This dataset contains 3D full-body gestures of five speakers, aligned speech audios, and Chinese text transcripts. The text transcripts were recognized by Alibaba Cloud Automatic Speech Recognition (ASR) service. The skeleton of this dataset is retargeted to be consistent with the Trinity Gesture dataset. To ensure semantic richness, speakers are instructed to cover a diverse set of topics, such as cooking, fiction, philosophy of life, and academic reporting. \fig\ref{fig:onset_intervals_Chinese} illustrates the distribution of onset intervals in our dataset, and \fig\ref{fig:lexicon_tsne_Chinese} shows the visualization of the learned gesture lexicon on our dataset.

\subsubsection{System Settings}
All the motion data are downsampled to $20$, $20$, and $15$ frames per second on the Trinity, Chinese, and TED datasets, respectively. The range of onset intervals $[D_m, D_M]$ is $[0.2s,$ $0.5s]$ for both the Trinity and TED datasets, but $[0.3s, 0.6s]$ for the Chinese dataset. The length of each normalized block $K$ $=$ $\lceil D_M\times{}fps \rceil$. The generator synthesizes $4$ seconds of gestures at a time. The dimensions of $d_t$, $d_{\tilde{a}}$, $d_{\tilde{A}}$, $d_{\tilde{m}}$, $d_s$, $d_z$, and $d_{P}$ are $768$, $128$, $128$, $128$, $192$, $32$, and $32$ respectively. The size of the gesture lexicon $N_s$ is $50$ for both the Trinity and Chinese datasets but $100$ for the TED dataset. We train our framework using the Adam optimizer with $\beta1 = 0.9$, $\beta2 = 0.999$ and a learning rate of $0.0003$. The loss weights $w_{\alpha}, w_{\beta}$, $w_{\eqword{rec}}$, $w_{\eqword{perc}}$, $w_{\eqword{lexeme}}$, and $w_{{z}}$ are set as $1.0$, $1.0$, $1.0$, $0.5$, $0.2$, and $1.0$, respectively. At runtime, we use a Gaussian filter with a kernel size of $K$ to smooth the denormalized gesture sequence, where $K=5$ is chosen to generate the results presented in this paper.

We train separate gesture generators on the Trinity, TED, and Chinese datasets. The cross-language capability of a generator can be further enhanced by pre-training the audio encoder (Section \ref{subsubsec:audio_representation}) using datasets in different languages. We have tried pre-training the audio encoder using both an English dataset (such as the Trinity or TED datasets) and our Chinese dataset and using it to train the generator on the Chinese dataset only. The gesture results can be found in the supplementary video.

\begin{figure*}[t]
    \centering
    \includegraphics[width=\textwidth]{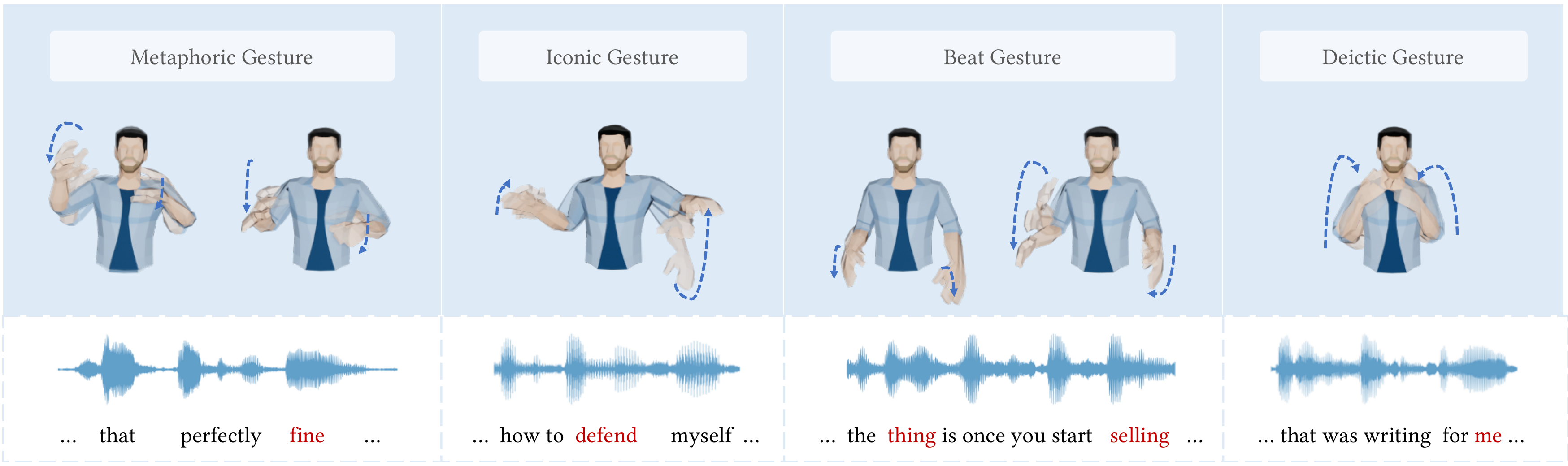}
    \caption{Qualitative results on the gestures synthesized by our method for four sample speech excerpts from the Trinity Gesture dataset \cite{ferstl2018investigating}. The character makes a metaphoric gesture when saying \emph{fine} and an iconic gesture for \emph{defend}. There are beat gestures for the words like \emph{thing} and \emph{selling}, and a deictic gesture appears when the character says \emph{me}. The drawing of this figure is inspired by \cite{yoon2020speech}.}
    \Description{}
    \label{fig:fig6}
\end{figure*}

\begin{figure*}[t]
    \centering
    \begin{subfigure}[t]{\textwidth}
        \centering
        \includegraphics[width=\textwidth]{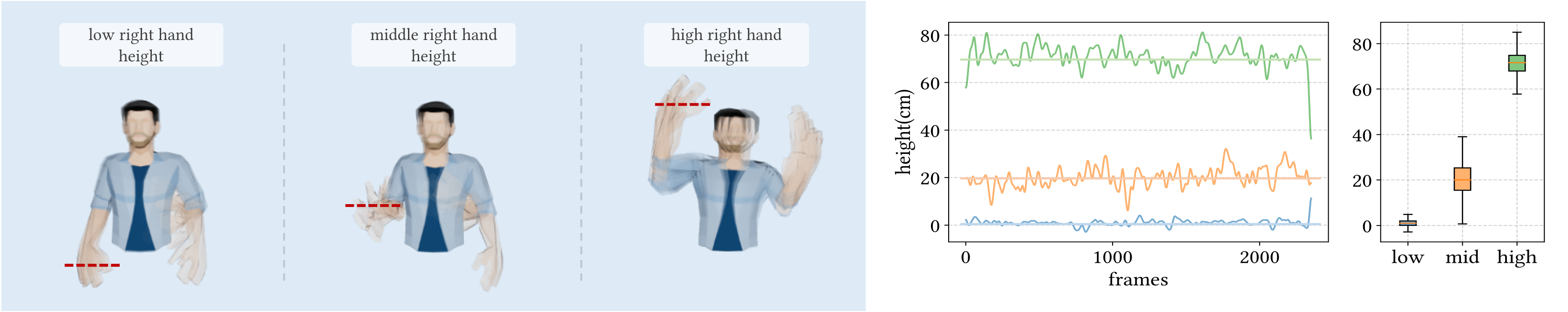}
        \caption{Low right hand, middle right hand, high right hand and time series statistics of right-hand height editing.}
        \label{fig:fig7a}
    \end{subfigure}
    \begin{subfigure}[t]{\textwidth}
        \centering
        \includegraphics[width=\textwidth]{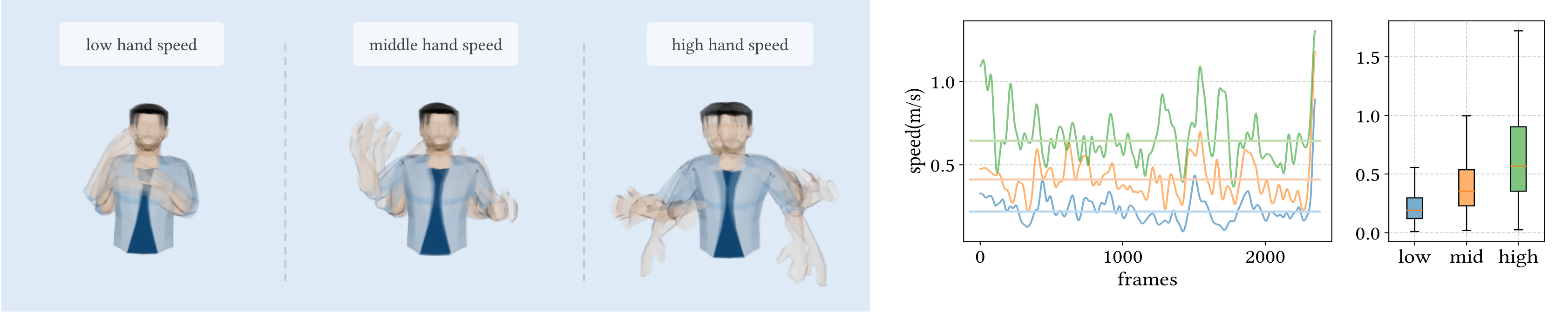}
        \caption{Low average speed, middle average speed, high average speed and time series statistics of speed editing.}
        \label{fig:fig7b}
    \end{subfigure}
    \begin{subfigure}[t]{\textwidth}
        \centering
        \includegraphics[width=\textwidth]{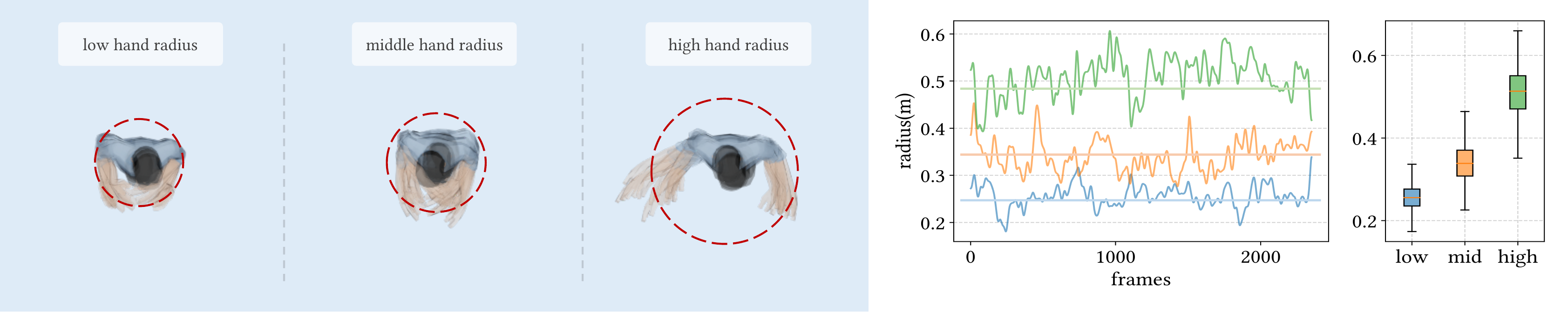}
        \caption{Low average radius, middle average radius, high average radius and time series statistics of radius editing.}
        \label{fig:fig7c}
    \end{subfigure}
    \caption{Results of style editing for the right-hand height (the first row), the hand speed (the second row), and the hand radius (the third row). The graphs on the right show the editing input (flat line) and the corresponding values of the output motions. The box plots show the statistics of the output.}
    \label{fig:fig7}
    \Description{}
\end{figure*}

\subsection{Evaluation}
\label{subsec:evaluation}
Figure \ref{fig:fig6} shows the gesture synthesis results for the speech excerpts from the test set of the Trinity dataset. Our system generates different types of realistic gestures. The character makes a metaphoric gesture when saying \emph{fine} and an iconic gesture for \emph{defend}. There are beat gestures for words like \emph{thing} and \emph{selling}, and a deictic gesture appears when the character says \emph{me}.

We also did a cross-language synthesis experiment to test the robustness of our system. We use the pre-trained model trained on the Trinity dataset (an English dataset) to generate gestures for a Chinese speech clip. Since different languages do not share the same word embedding, we generate gestures by taking only the speech audio as input (Section \ref{subsec:inference_based_on_audio_only}). As illustrated in Figure \ref{fig:fig9b}, when encountering a different language, our system still generates beat-matched and stylized gestures, reflecting the robustness of our system. We also trained our system on the Chinese dataset and then did another cross-language experiment by generating gestures for an English speech excerpt. Please refer to the supplementary video for the visualization results.

\subsubsection{Style Editing}
\label{subsubsec:style_editing}
Inspired by the idea of \citet{alexanderson2020style}, we augment our system to achieve motion editing by adding a feature block $\vect{C} \in\mathbb{R}^{K\times{}d_{c}}$ of the gesture motion as an extra input of the generator and the lexeme interpreter (\fig\ref{fig:interpreters}). The computations in \eqn\eqref{eqn:generator_block} and \eqref{eqn:lex_interpreter} are then reformulated as
\begin{align}
    \vect{M}_{i}^* &=\mathcal{G}(\tildevect{M}_{i-1},\tildevect{A}_{i},\vect{s}_i,\vect{z}_i,\vect{P},\vect{C}_i), \\
    \vect{s}_i^* &=\mathcal{P}_s(\vect{s}_{i-1}^*, \tildevect{A}_i^{\eqword{H}}, \tildevect{T}_i, \vect{I},\vect{C}_i), 
\end{align}
which allows the network to learn the relationship between a desired motion feature and a gesture motion. During inference phase, we can easily edit the motion style feature of the generated gestures by modifying the feature block $\vect{C}$ as needed.

Similar to \cite{alexanderson2020style}, we have experimented with three different style features using our system: the height of the right-hand (hand height), the average speed of both hands (hand speed), and the average distance from the hand positions to the up-axis through the root joint of the character (hand radius). We train a separate generator for each style feature. The training data is computed using the reference motions and averaged within a four-second sliding window, forming $K$-frame feature blocks. 

We have synthesized three animations for each of the motion styles. Each animation has a constant desired low, mid, or high feature value, as shown in the first three columns of \fig\ref{fig:fig7}. The last column of \fig\ref{fig:fig7} shows the accuracy of the generated motion features. These results indicate that all the editing signals could efficiently affect the generated gestures.

\subsection{Comparisons}
\label{sbusec:comparisons}
In this section, we compare our system with several state-of-the-art systems to demonstrate the advances made by our system. We first briefly introduce these systems and then our quantitative and qualitative comparisons. We also propose a simple but effective objective metric (PMB) to evaluate the rhythm performance for co-speech gesture synthesis.

\begin{table*}[t]
    \centering
    \caption{Comparison of our system to SG \cite{alexanderson2020style}, Ges \cite{kucherenko2020gesticulator}, GTC \cite{yoon2020speech}, and S2AG \cite{bhattacharya2021speech2affectivegestures} on the TED and Trinity datasets. The system without beat segmentation (w/o BC) uses a fixed interval of $D_M$ for segmentation, which is $0.5s\sim{}0.6s$ depending on which dataset is used. The system without gesture lexeme (w/o SC) excludes the gesture lexicon and lexeme interpreter modules. The generator is retrained to predict future gestures based on only the previous motion, the audio, and the style code. Similarly, the system without gesture style code (w/o ZC) excludes the style code and the style interpreter modules. Only the motion, the audio, and the lexeme are used by the generator. \emph{Ours (audio only)} denotes the audio-only inference.}
    \label{tab:table1}
    
    \newcolumntype{Y}{>{\raggedleft\arraybackslash}X}
    \newcolumntype{Z}{>{\centering\arraybackslash}X}
    \begin{tabularx}{\linewidth}{XXYYYY}
        \toprule
        Dataset & System & MAJE ($mm$) $\downarrow$ & MAD ($mm/s^2$) $\downarrow$ & FGD $\downarrow$ & PMB ($\%$) $\uparrow$ \\
        \toprule
        \multirow{8}*{Trinity} & Real Gesture & 0.0 & 0.0 & - & 95.74 \\
        \cline{2-6}
        & SG & 97.29 & 4.26 & 36.98 & 54.54 \\
        & Ges & 82.41 & 3.62 & 31.04 & 71.0 \\
        & S2AG & 54.93 & 1.49 & 20.36 & 79.53 \\
        \cline{2-6}
        & Ours (w/o BC) & 59.11 & 1.89 & 16.13 & 78.18 \\
        & Ours (w/o SC) & 70.10 & 2.51 & 29.75 & 85.74 \\
        & Ours (w/o ZC) & 52.85 & 1.35 & 12.53 & $\bm{91.36}$ \\
        & Ours (audio only) & 57.99 & 1.83 & 15.79 & 87.35 \\
        & Ours & $\bm{49.53}$ & $\bm{0.97}$ & $\bm{10.78}$ & $\bm{91.36}$ \\
        
        \midrule
        \multirow{7}*{TED} & Real Gesture & 0.0 & 0.0 & - & 93.10 \\
        \cline{2-6}
        & GTC & 26.95 & 3.03 & 3.73 & 71.72 \\
        & S2AG & 24.49 & 2.93 & 3.54 & 75.57 \\
        \cline{2-6}
        & Ours (w/o BC) & 27.10 & 3.11 & 3.88 & 67.88 \\
        & Ours (w/o SC) & 30.07 & 3.53 & 5.22 & 83.10 \\
        & Ours (w/o ZC) & 21.33 & 2.61 & 2.47 & 88.67 \\
        & Ours (audio only) & 27.28 & 3.17 & 3.96 & 81.33 \\
        & Ours & $\bm{18.13}$ & $\bm{2.29}$ & $\bm{2.04}$ & $\bm{89.52}$ \\
        \bottomrule
    \end{tabularx}
    
\end{table*}

\begin{figure}[t]
    \centering
    \begin{subfigure}[t]{0.47\linewidth}
        \centering
        \includegraphics[width=\linewidth]{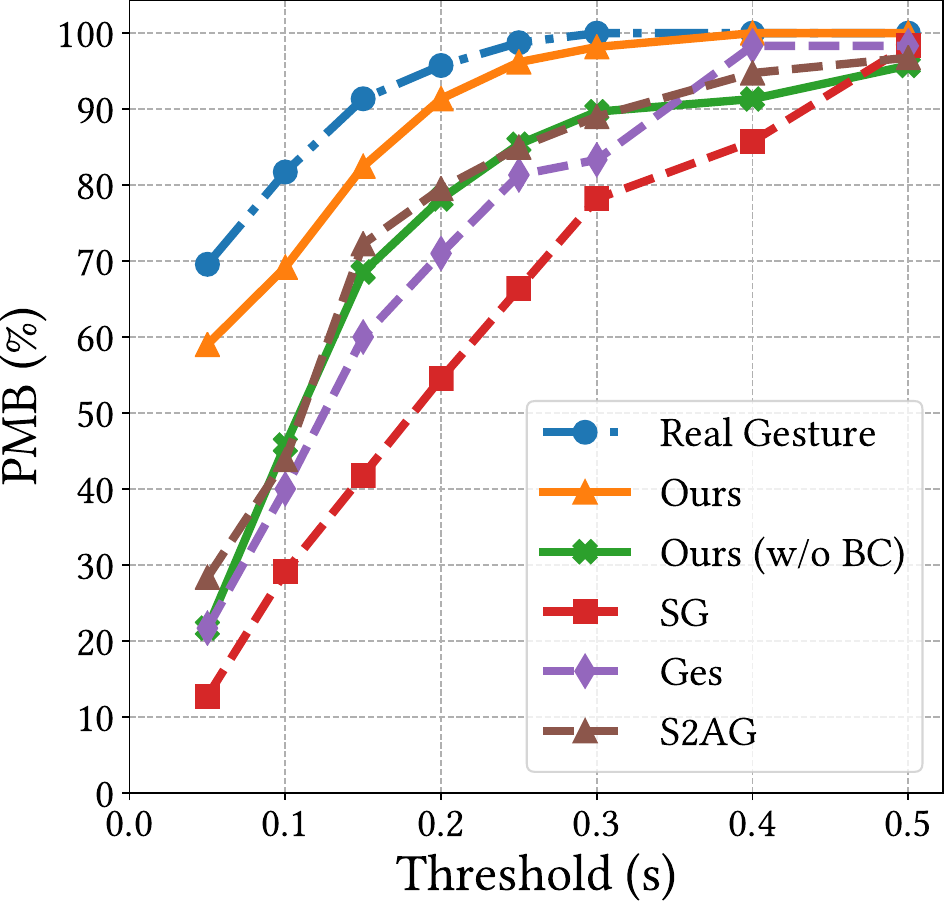}
        \caption{Trinity Gesture Dataset}
        \label{fig:table1a}
    \end{subfigure}
    \hspace{\fill}
    \begin{subfigure}[t]{0.47\linewidth}
        \centering
        \includegraphics[width=\linewidth]{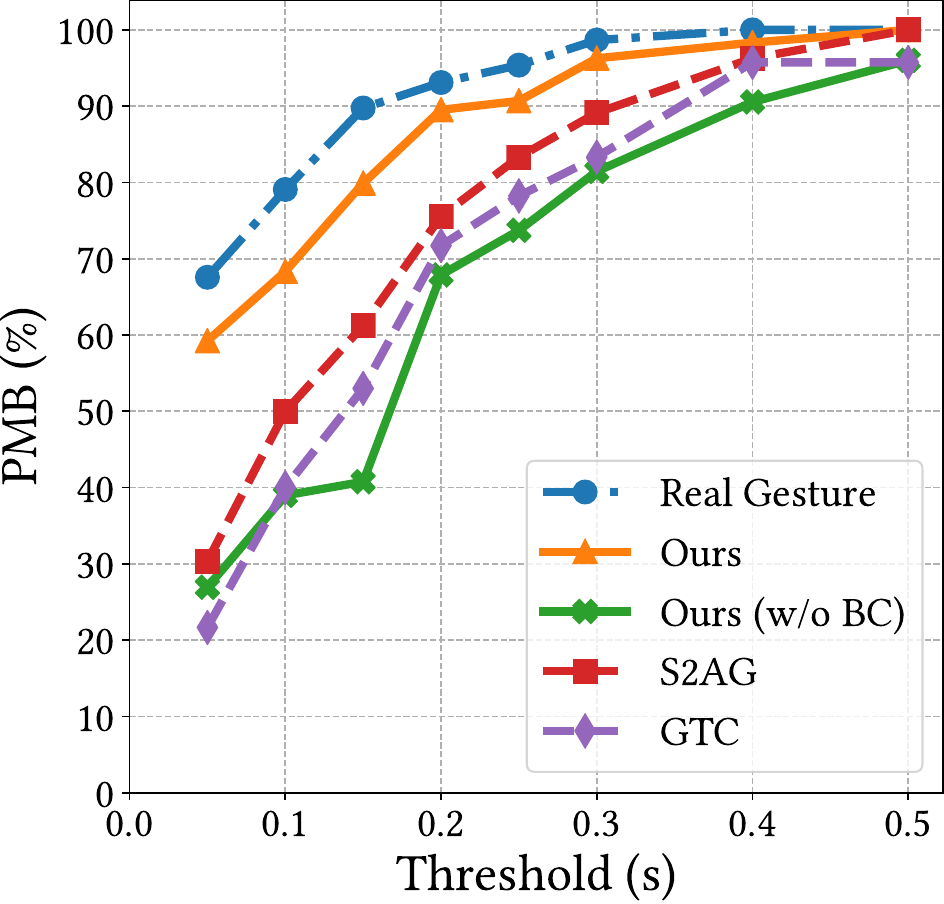}
        \caption{TED Gesture dataset}
        \label{fig:table1b}
    \end{subfigure}
    \Description{}
    \caption{PMB results of continuously adjusting the threshold $\delta$ on the Trinity and TED datasets.}
    \label{fig:table1}
\end{figure}

\subsubsection{Systems of Comparison}
We choose four recent successful 3D co-speech gesture synthesis systems for comparison. On the Trinity dataset, we compare with the systems of Style Gesture (SG) \cite{alexanderson2020style} and Gesticulator (Ges) \cite{kucherenko2020gesticulator}. SG generates gestures based on only the speech audio and uses a normalizing flow model. Ges leverages the audio and text of the speech to generate semantically consistent gestures. On the TED dataset, we compare the systems of Gestures from Trimodal Context (GTC) \cite{yoon2020speech} and Speech to Affective Gestures (S2AG) \cite{bhattacharya2021speech2affectivegestures}. GTC uses speech audio, text transcript, and speaker identity to generate gestures. Based on the three modalities used in GTC, S2AG adds another new modality of affective expressions from the seed poses into their model.

Because we have no access to the official pretrained model of SG, we strictly follow the official configuration and run the codes offered by authors to train a model. For other systems, we use the pretrained models provided by the authors. For a fair comparison, we use the same skeleton and motion frame rate as the baselines.

\subsubsection{Quantitative Evaluation}

\begin{figure*}[t]
    \centering
    \begin{subfigure}[t]{0.48\textwidth}
        \centering
        \includegraphics[width=\textwidth]{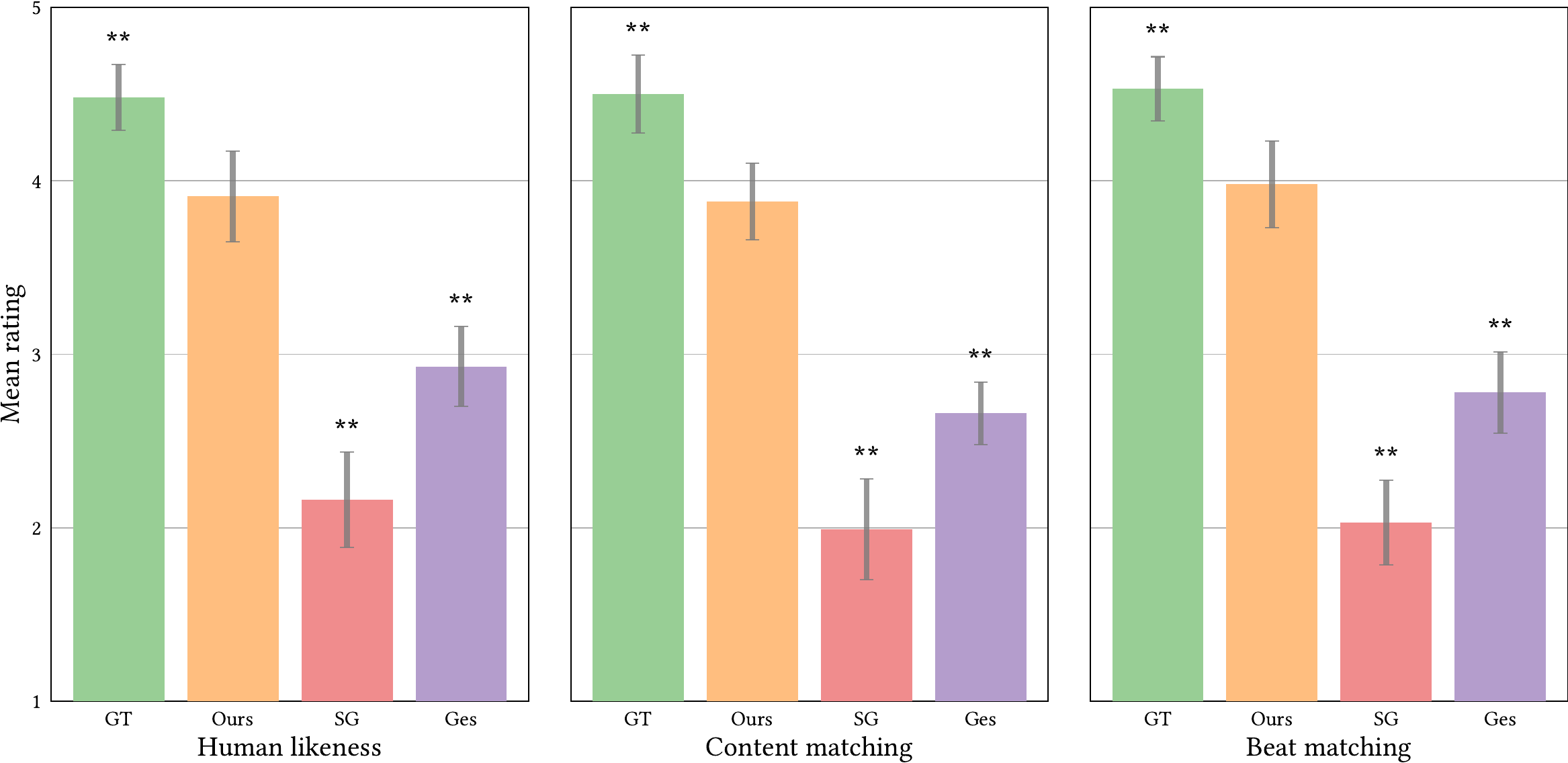}
        \caption{Mean ratings for the English speech clips.}
        \label{fig:fig8a}
    \end{subfigure}
    \hspace{\fill}
    \begin{subfigure}[t]{0.48\textwidth}
        \centering
        \includegraphics[width=\textwidth]{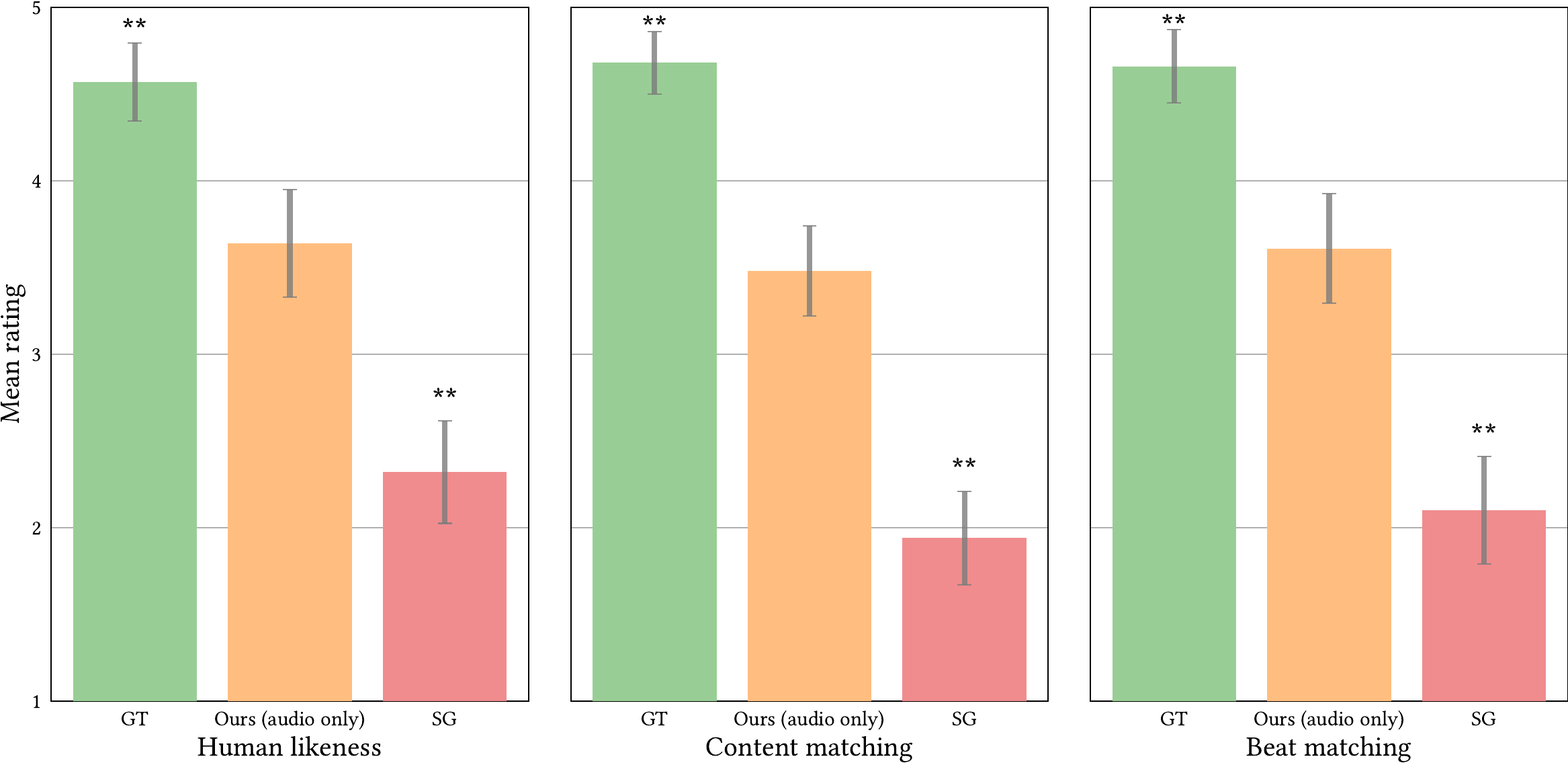}
        \caption{Mean ratings for the Chinese speech clips.}
        \label{fig:fig8b}
    \end{subfigure}
    \caption{User study results with $95\%$ confidence intervals. Asterisks indicate the significant effects ($*: p < 0.05$, $**: p < 0.01$, $***: p < 0.001$). All the models are trained on the Trinity dataset (an English dataset). See Section \ref{subsubsec:user_study} for details.}
    \Description{}
    \label{fig:fig8}
\end{figure*}

We first adopt three commonly used evaluation metrics (MAJE, MAD, and FGD) to compare these systems quantitatively. 
MAJE measures the \textbf{m}ean of the \textbf{a}bsolute \textbf{e}rrors between the generated \textbf{j}oint positions and the ground truth over all the time steps and joints, which indicates how closely the generated joint positions follow the ground truth. 
MAD measures the \textbf{m}ean of the $\ell_2$ norm \textbf{d}ifferences between the generated joint \textbf{a}ccelerations and the ground truth over all the time steps and joints, which indicates how closely the ground truth and the generated joint movements match.
\textbf{F}r{\'e}chet \textbf{G}esture \textbf{D}istance (FGD) was proposed by \citet{yoon2020speech}, which measures the difference between the distributions of the latent features of the generated gestures and ground truth, where the latent features are extracted using an auto-encoder trained on the Human 3.6M dataset \cite{ionescu2013human3}. FGD could assess the perceived plausibility of the synthesized gestures.  

Calculating the matching rate between audio and motion beats is a standard method of evaluating rhythm performance, which has been widely used in music-driven dance synthesis \cite{Li_2021_aist, chen2021choreomaster}. However, we cannot simply apply this method to the gesture generation task because the onset of beat gesture usually precedes the corresponding speech beat by a small amount of time \cite{pouw2019quantifying}. Thus, we need a distance threshold to determine robustly whether the audio and motion beats match each other. 

\textbf{P}ercentage of \textbf{c}orrectly predicted \textbf{k}eypoints (PCK)  is a widely used metric in human pose estimation \cite{Wei_2016_openpose, mehta2017vnect}, where a predicted key point is considered correct if its distance to the ground truth is smaller than an adjustable threshold. Inspired by this metric, we propose a new metric, \emph{PMB}, as the \textbf{p}ercentage of \textbf{m}atched \textbf{b}eats, where a motion beat is considered to be matched if its temporal distance to a nearby audio beat is smaller than a threshold $\delta$. Specifically,
\begin{equation}
    \eqword{PMB}(\vect{B}^{{m}}, \vect{B}^{{a}}) = \frac{1}{N_m}\sum^{N_m}_{i=1}\sum^{N_a}_{j=j_{[i-1]}^{{a}*}+1} \mathbbm{1}\big[\lVert \vect{b}_i^{{m}} - \vect{b}_j^{{a}} \rVert_1 < \delta\big].
\end{equation}
The sequence of motion beats $\vect{B}^{{m}}=\{\vect{b}_1^{{m}}, \dots, \vect{b}_{N_m}^{{m}}\}$, where $N_m$ is the number of motion beats, is identified using the algorithm proposed by \citet{ho2013motionbeat} based on the local minima of joint deceleration. 
The audio beats $\vect{B}^{{a}}=\{\vect{b}_1^{{a}}, \dots, \vect{b}_{N_a}^{{a}}\}$ are the onset sequence, where $N_a$ is the number of onsets. $j_{[i-1]}^{{a}*}$ represents the index of the audio beats that the last motion beat matches. $\mathbbm{1}$ denotes the indicator function. The threshold $\delta$ is set to $0.2$s by default, but we can adjust $\delta$ continuously and observe the changes of the PMB values, which provides a more comprehensive picture of the rhythmic performance.

Table \ref{tab:table1} summarizes the performance of all the systems on the two English datasets, Trinity and TED. For the MAJE, MAD, and FGD metrics, our system achieves the lowest values on both datasets. Note that the FGD values of our system are significantly lower than other systems, which indicates the better perceptual quality of gestures synthesized by our system. It is interesting that the FGD values rise rapidly without the gesture lexeme component (w/o SC), which means the gesture lexeme is crucial to gesture quality. There is a decline in the generation performance in the audio-only inference because of the lack of sufficient input information. However, the generated gestures are still acceptable.

As for the rhythm performance, our system achieves the highest PMB values on both datasets. The PMB values drop rapidly without the beat segmentation component (w/o BC), indicating that beat segmentation is vital to rhythm awareness. Figure \ref{fig:table1} shows the PMB results with different thresholds. It can be seen that even the ground-truth motion does not match the speech beats precisely, while the behavior of our method is closer to the ground truth compared  with the baselines.

\begin{figure*}[t]
    \centering
    \begin{subfigure}[t]{0.48\textwidth}
        \centering
        \includegraphics[width=\textwidth]{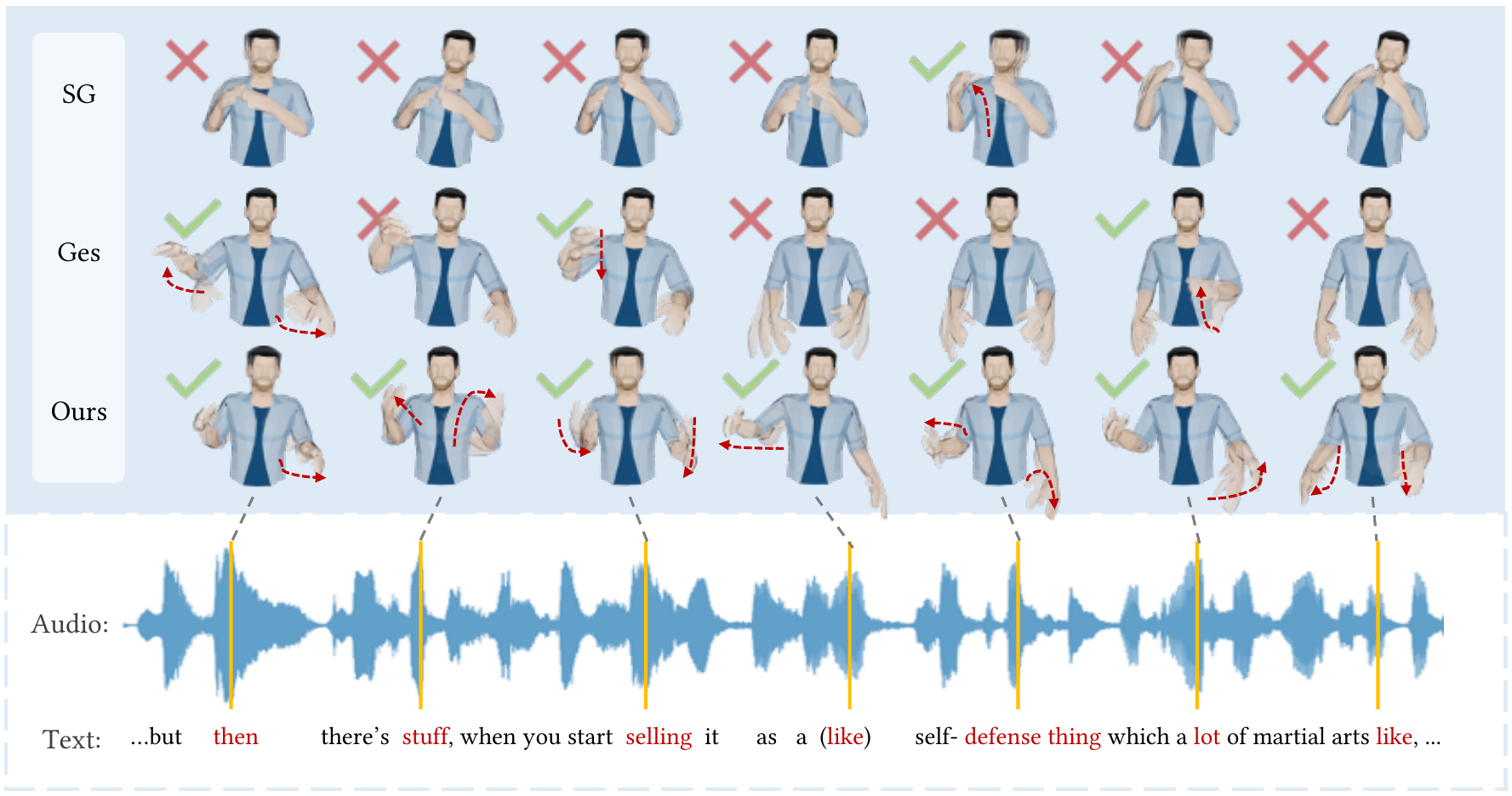}
        \caption{Gesture results for an English speech clip.}
        \label{fig:fig9a}
    \end{subfigure}
    \hspace{\fill}
    \begin{subfigure}[t]{0.48\textwidth}
        \centering
        \includegraphics[width=\textwidth]{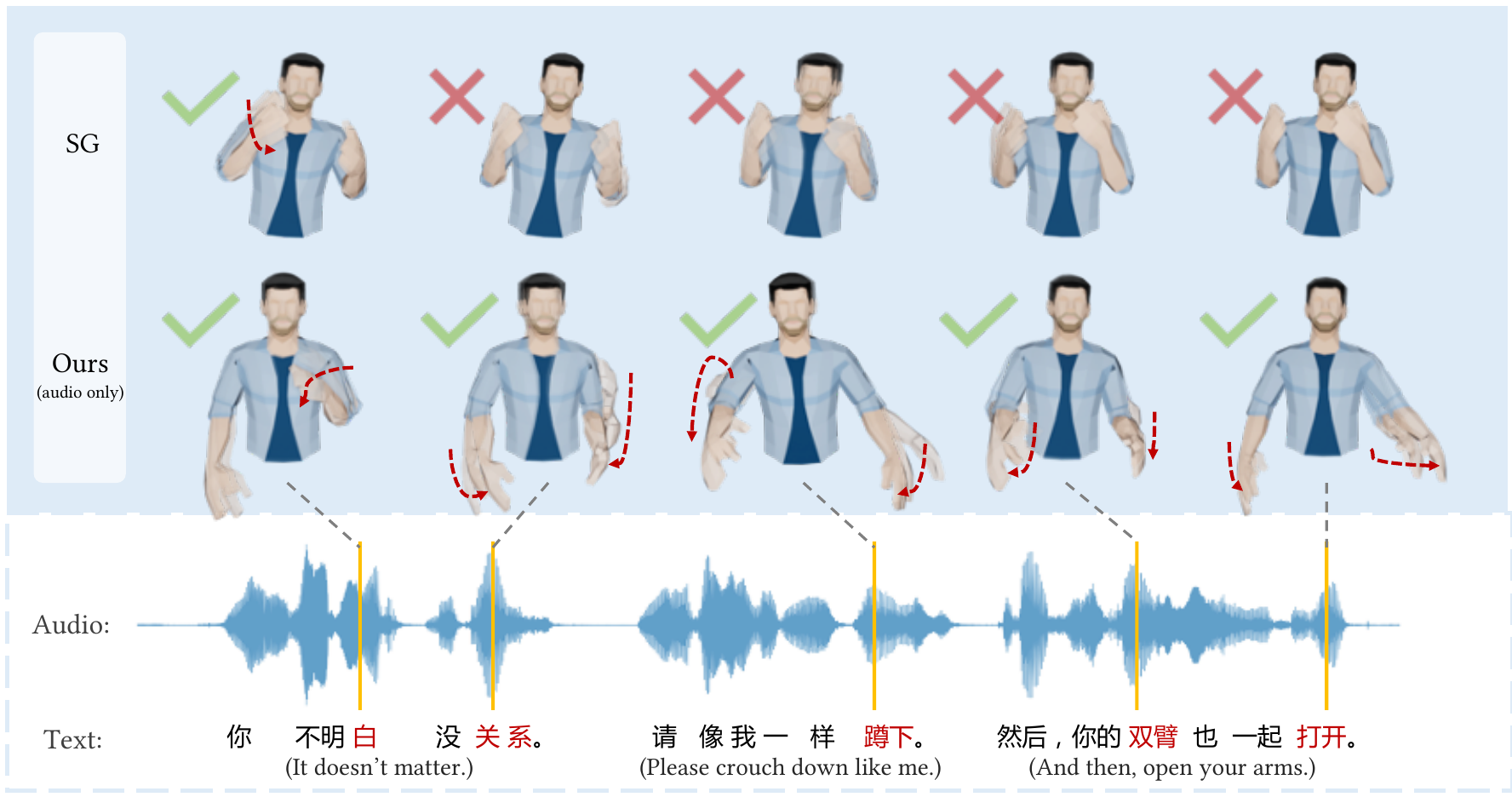}
        \caption{Gesture results for a Chinese speech clip.}
        \label{fig:fig9b}
    \end{subfigure}
    \caption{Generated motions of SG, Ges, and our system for the same input speech used in the user study. All the models are trained on the Trinity dataset (an English dataset). The red words indicate beats. The red arrows show the movement of corresponding beat gestures. A green check indicates a correct beat match, while a red cross indicates a wrong beat match.}
    \label{fig:fig9}
    \Description{}
\end{figure*}

\subsubsection{User Study}
\label{subsubsec:user_study}
We further conduct user studies to assess the performance of our system qualitatively, where SG \cite{alexanderson2020style} and Ges \cite{kucherenko2020gesticulator} are used as the baseline methods. 
We generate $14$ test video clips, each consisting of the results synthesized by our methods, the baselines, and the ground truth in random order. Among the $14$ clips, nine are English clips generated from the test set of the Trinity dataset, and five are Chinese clips generated from the test set of our Chinese dataset. Notably, all the clips are synthesized using the models trained on the English Trinity dataset. The duration of each clip is around $30$s. 

We have recruited $30$ volunteer participants to participate in our study, of which $19$ are male and $11$ are female. 19 participants are $18-22$ years of age, 10 are between $23-30$, and one is above $30$. When participants watch each video clip, they will be asked to answer three questions and rate the video from $1$ to $5$, with $1$ being the worst, $3$ being average, and 5 being the best. The three questions are: (1) human likeness (neglecting the speech audio), (2) speech-to-gesture content matching, and (3) speech-to-gesture beat matching. 

The results of these user studies are shown in \fig\ref{fig:fig8}, our system receives higher scores than the other systems and is closer to the real gestures (GT). A one-way ANOVA reveals main effects of \emph{human likeness}, \emph{content matching}, and \emph{beat matching}, and a post-hoc Tukey multiple comparison test identifies a significant difference ($p < 0.005$) between our system and all the other methods. 
As also illustrated in \fig\ref{fig:fig9a}, the end-to-end systems SG and Ges are less sensitive to rhythm than ours, and the resulting motions lack diversity, which affects their performance in the user studies.

The statistical results of the cross-language test (\fig\ref{fig:fig8b}) demonstrate the better robustness of our system. When dealing with a completely different language (\fig\ref{fig:fig9b}), the gestures generated by SG are more rigid and do not match the beats correctly. In contrast, our model (audio-only) is still able to perceive beats accurately and generate dynamic gestures. Notably, we do not compare with Ges in this cross-language test because this model only supports English text transcripts. 

\subsection{Ablation Study}
\label{subsec:ablation_study}
We conduct a variety of ablation studies to analyze the performance of our system. Notably, only the ablation of the lexicon size (\fig\ref{fig:fig12}) uses the validation set of the dataset to determine the hyperparameter of the model. All other experiments are based on the test set.

\subsubsection{Hierarchical Audio Feature Disentanglement}
\label{subsubsec:disentanglement}
\begin{figure}[t]
    \centering
    \begin{subfigure}[t]{0.62\linewidth}
        \centering
        \includegraphics[width=\linewidth]{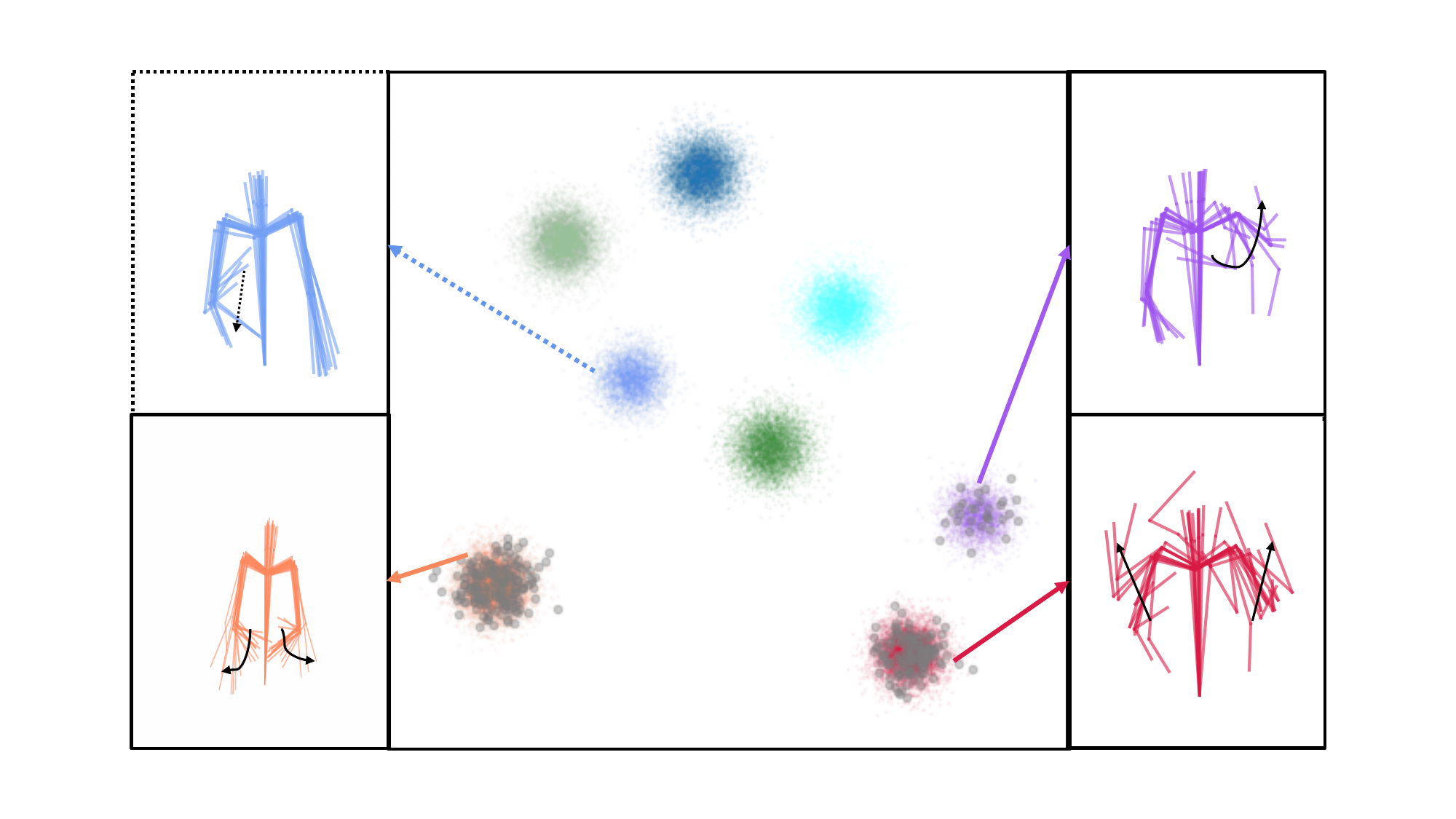}
        \caption{High-level}
        \label{fig:fig14a}
    \end{subfigure} 
    \hspace{\fill}
    \begin{subfigure}[t]{0.3554\linewidth}
        \centering
        \includegraphics[width=\linewidth]{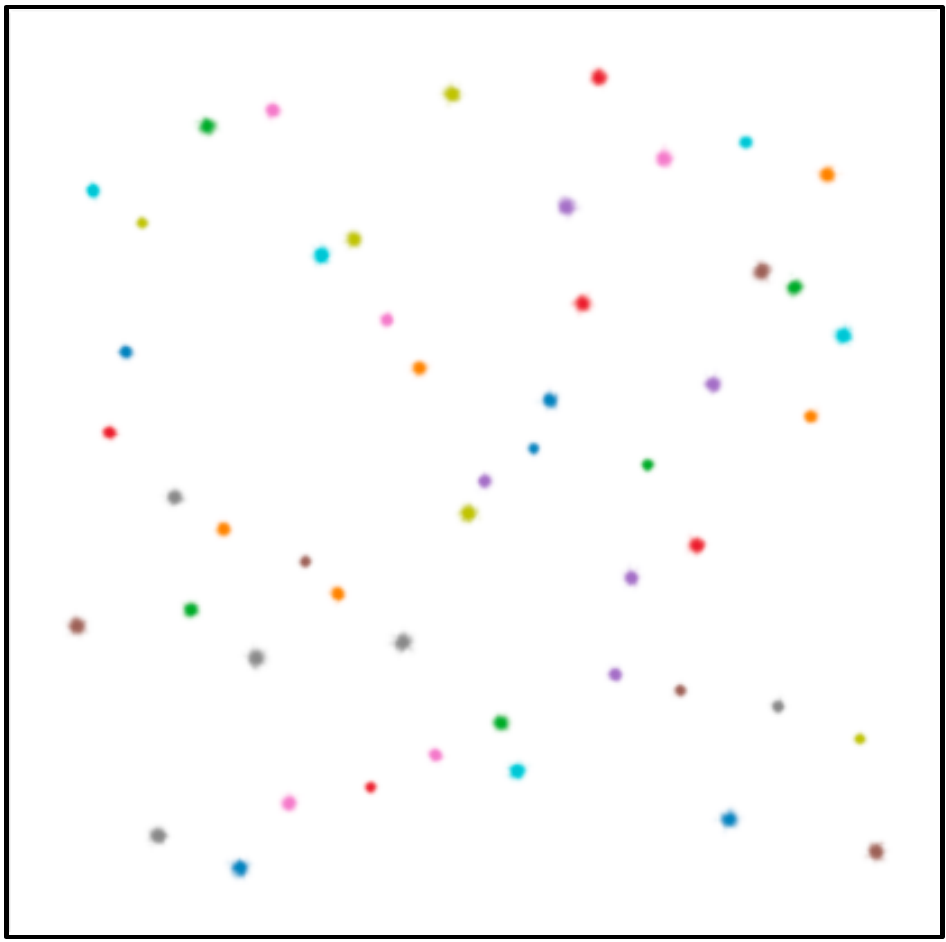}
        \caption{Low-level}
        \label{fig:fig14b}
    \end{subfigure} 
    \caption{t-SNE visualization of motion latent codes. Each color (except gray) stands for a gesture lexeme. 
    (a) latent codes (gray dots) corresponding to cluster $\vect{C}^{\eqword{high}}_0$ only appear in specific gesture lexemes (orange, purple, and red).
    (b) latent codes (various color dots) corresponding to cluster $\vect{C}^{\eqword{low}}_0$ are distributed in most of the gesture lexemes.
    See Section \ref{subsubsec:disentanglement} for details.}
    \label{fig:fig14}
    \Description{}
\end{figure}

We presume that the high-level audio feature $\vect{a}^{\eqword{high}}$ contains semantics-relevant information that determines the gesture lexeme $\vect{s}$.
To justify this assumption, we apply the K-means algorithm to all the high-level audio blocks of the TED Gesture dataset and get 50 clusters, where each cluster essentially indexes the audio clips with similar semantics. We can find several representative clusters $\vect{C}^{\eqword{high}}_0$ whose corresponding text transcripts contain words with a similar meaning, such as \emph{many}, \emph{quite a few}, \emph{lots of}, \emph{much}, and \emph{more}, etc. Meanwhile, these audio clips also correspond to a set of generated motion blocks $\{\vect{M}^*_0,\vect{M}^*_1,\dots\}$. By encoding these motion blocks using the pre-trained encoder $\mathcal{E}$, we can obtain their corresponding motion latent codes. As illustrated in \fig\ref{fig:fig14a}, these latent codes (gray dots) only appear in a few gesture lexemes (orange, purple, and red), and it can be seen that the sample gestures of these latent codes indeed convey the semantics of the cluster $\vect{C}^{\eqword{high}}_0$.
The same observation is not true for the low-level audio features. If we also cluster all low-level audio features and pick a representative cluster $\vect{C}^{\eqword{low}}_0$, \fig\ref{fig:fig14b} shows that the corresponding motion latent codes (various color dots) appear nearly uniformly in most of the gesture lexemes.
The experiments above confirm the correlation between the high-level audio features and the gesture lexemes, as well as the semantic disentanglement between high-level and low-level audio features.

\subsubsection{Gesture Style Code}
\label{subsubsec:gesture_style_code}
\begin{figure}[t]
    \centering
    \begin{subfigure}[t]{0.32\linewidth}
        \centering
        \includegraphics[width=\linewidth]{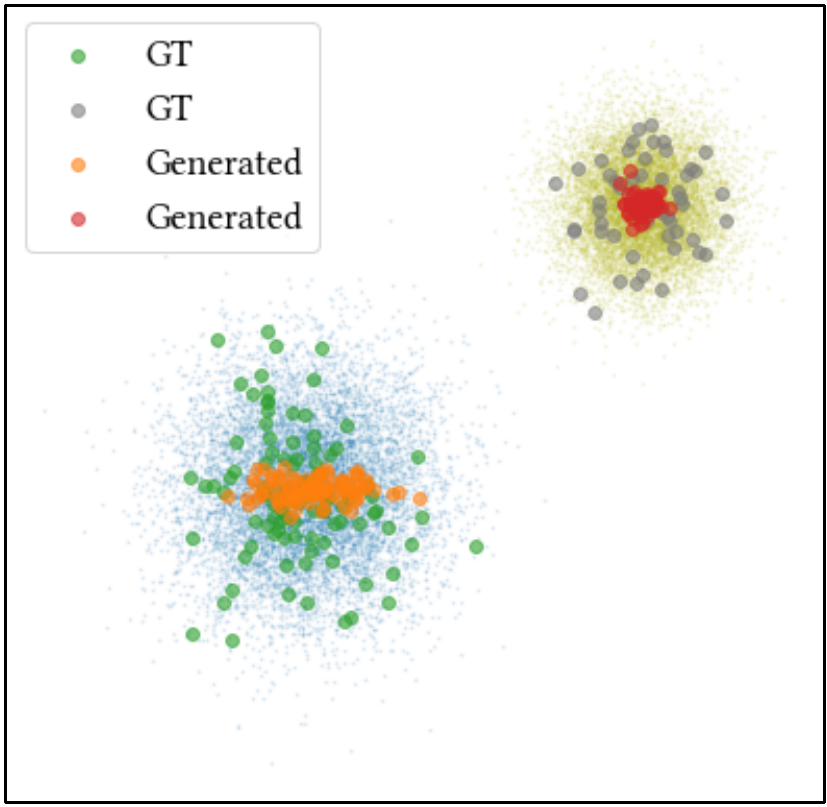}
        \caption{w/o $\vect{a}^{\eqword{low}},\vect{z}$}
        \label{fig:fig15a}
    \end{subfigure} 
    \hspace{\fill}
    \begin{subfigure}[t]{0.32\linewidth}
        \centering
        \includegraphics[width=\linewidth]{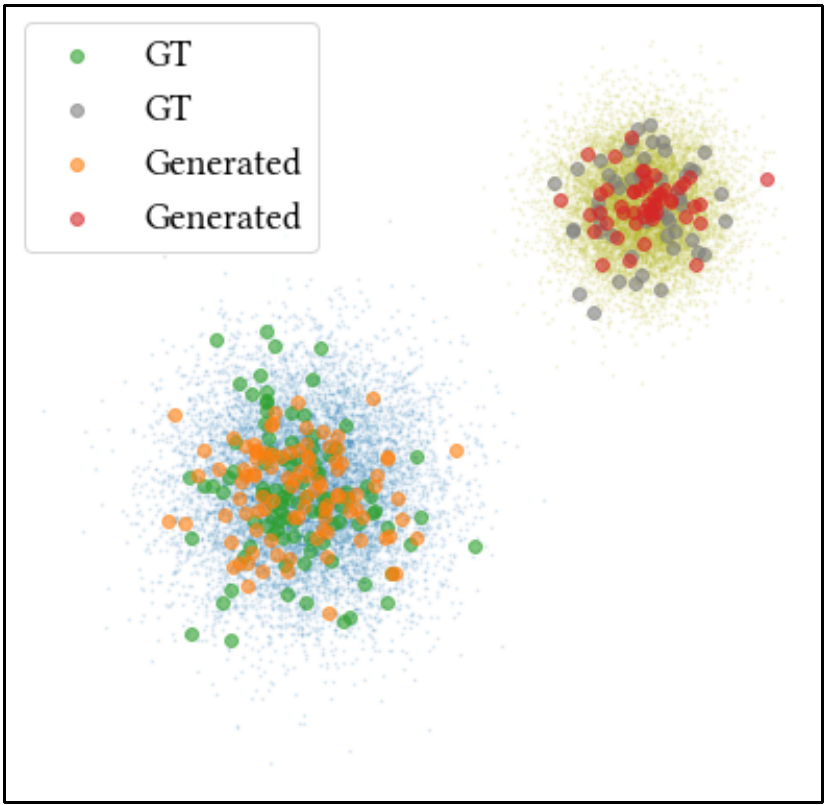}
        \caption{w/ $\vect{a}^{\eqword{low}}$ + w/o $\vect{z}$}
        \label{fig:fig15b}
    \end{subfigure} 
    \hspace{\fill}
    \begin{subfigure}[t]{0.32\linewidth}
        \centering
        \includegraphics[width=\linewidth]{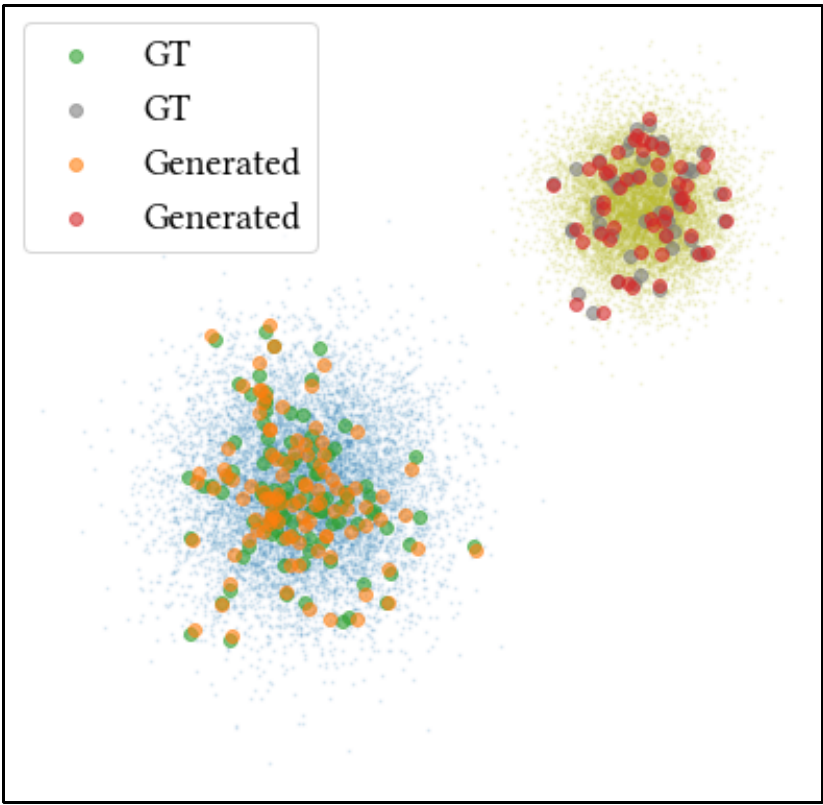}
        \caption{w/ $\vect{a}^{\eqword{low}},\vect{z}$}
        \label{fig:fig15c}
    \end{subfigure} 
    \caption{t-SNE visualization of real gestures (GT) versus generated gestures in the motion latent space. We randomly choose two gesture lemexes for visualization.
    (a) results of our system without low-level audio feature and gesture style code.
    (b) with low-level audio feature but without gesture style code.
    (c) with low-level audio feature and gesture style code.
    See Section \ref{subsubsec:disentanglement} for details.}
    \label{fig:fig15}
    \Description{}
\end{figure}
\begin{table}[t]
    \centering
    \caption{Comparison of style interpreters w/ and w/o low-level audio features and gesture style code. See Section \ref{subsubsec:disentanglement} for details.}
    \label{tab:table6}
    
    \begin{tabularx}{\linewidth}{lXrrr}
        \toprule
        Dataset & System & Variance $\uparrow$ & FGD $\downarrow$ & PMB ($\%$) $\uparrow$ \\
        \toprule
        \multirow{5}*{Trinity} & Real Gesture & 0.41 & - & 95.74 \\
        \cline{2-5}
        & w/o $\vect{a}^{\eqword{low}},\vect{z}$ & 0.09 & 17.99 & 89.35 \\
        & w/ $\vect{a}^{\eqword{low}}$ + w/o $\vect{z}$ & 0.21 & 12.53 & $\bm{91.36}$ \\
        & w/ $\vect{a}^{\eqword{low}},\vect{z}^{\eqword{random}}$ & 0.30 & 11.96 & 91.28 \\
        & w/ $\vect{a}^{\eqword{low}},\vect{z}$ & $\bm{0.37}$ & $\bm{10.78}$ & $\bm{91.36}$ \\
        
        \midrule
        \multirow{5}*{TED} & Real Gesture & 2.72 & - & 93.10 \\
        \cline{2-5}
        & w/o $\vect{a}^{\eqword{low}},\vect{z}$ & 0.89 & 3.71 & 84.57 \\
        & w/ $\vect{a}^{\eqword{low}}$ + w/o $\vect{z}$ & 1.87 & 2.47 & 88.67 \\
        & w/ $\vect{a}^{\eqword{low}},\vect{z}^{\eqword{random}}$ & $\bm{2.49}$ & 2.13 & 88.89 \\
        & w/ $\vect{a}^{\eqword{low}},\vect{z}$ & 2.45 & $\bm{2.04}$ & $\bm{89.52}$ \\
        \bottomrule
    \end{tabularx}
  
\end{table}
\begin{table}[t]
    \centering
    \caption{Comparison of style interpreters w/ and w/o text features.}
    \label{tab:table5}
    
    \newcolumntype{Y}{>{\raggedleft\arraybackslash}X}
    \newcolumntype{Z}{>{\centering\arraybackslash}X}
    \begin{tabularx}{\linewidth}{XlYY}
        \toprule
        Dataset & System & FGD $\downarrow$ & PMB ($\%$) $\uparrow$ \\
        \toprule
        \multirow{2}*{Trinity} & w/o text feature & 10.91 & $\bm{91.36}$ \\
        & w/ text feature & $\bm{10.78}$ & $\bm{91.36}$ \\
        
        \midrule
        \multirow{2}*{TED} & w/o text feature & 2.09 & 89.22 \\
        & w/ text feature & $\bm{2.04}$ & $\bm{89.52}$ \\
        \bottomrule
    \end{tabularx}
\end{table}
The low-level audio feature $\vect{a}^{\eqword{low}}$ contains semantic-irrelevant information, e.g., pitch and volume. Presumably, it should affect the motion variations within a gesture lexeme. In our system, we introduce the learnable gesture style code $\vect{z}$ combined with the low-level audio feature to jointly determine the motion variations, which can be considered a fine-grained style.
As illustrated in Figure \ref{fig:fig15b}, the low-level audio feature increases the variety of the generated gestures but cannot fully decide the variations. With the introduction of the gesture style code (Figure \ref{fig:fig15c}), the distribution of generated gestures becomes closer to the ground-truth distribution. 

Table \ref{tab:table6} compares several settings in terms of motion variance, FGD, and PMB. We measure the Euclidean distance from the motion latent code of a motion to the corresponding gesture lexeme and compute the variance of these distances corresponding to the same lexeme. The \emph{motion variance} is then defined as the average of such variances of every lexeme. $\vect{z}$ denotes the output of the style interpreter, while $\vect{z}^{\eqword{random}}$ is a random style code sampled from the normal distribution in the latent space. Consistent with Figure \ref{fig:fig15}, Table \ref{tab:table6} also shows that combining the low-level audio feature and styles achieves more significant motion variance and lower FGD values. Besides, the FGD values in Table \ref{tab:table6} also indicate that the style codes computed by the interpreter generate gestures that are more perceptually acceptable than that created using random style codes.

We have also evaluated the importance of the text features in interpreting the gesture style codes (see \eqn\ref{eqn:style_interpreter}).
The result of Table \ref{tab:table5} shows that interpreting the style codes 
with text features does improve the FGD value. However, the improvement is marginal. Considering the inference efficiency, we can interpret the style codes conditional on only the low-level audio features.

\subsubsection{Range of Onset Intervals}
\label{subsubsec:onset_interval}
Choosing a proper range of onset intervals is crucial to achieving quality gestures. Intuitively, the lower bound of this range affects the model's sensitivity to beat. If the lower bound is too high (such as 0.5s), the interval between the identified beats becomes large, causing the model to respond sluggishly to beats. The upper bound of the onset intervals regularizes the variance of the duration of the speech blocks, which should not be too large either. If the length of the speech blocks differs too much, the side effect of the normalization becomes visible, causing unnatural transitions between the generated gesture blocks. 

To verify the above hypothesis, we compare the performance of three different interval ranges: 0.2-0.5s, 0.2-1.0s, and 0.5-1.0s, where 0.2-0.5s is our default setting (Section \ref{subsubsec:onset_identification}).
As shown in Table \ref{tab:table2}, our default range of onset intervals achieves the best FGD and PMB values on both the Trinity and TED Gesture datasets. PMB drops significantly when the minimum distance between onsets is high (0.5-1.0s), which indicates that the model is insensitive to the rhythm. When the variance is too large (0.2-1.0s), PMB is barely affected, but FGD drops a lot, showing that the generated motions exhibit a lower quality. This result is consistent with our hypothesis.

\begin{figure}[t]
    \centering
    \includegraphics[width=\linewidth]{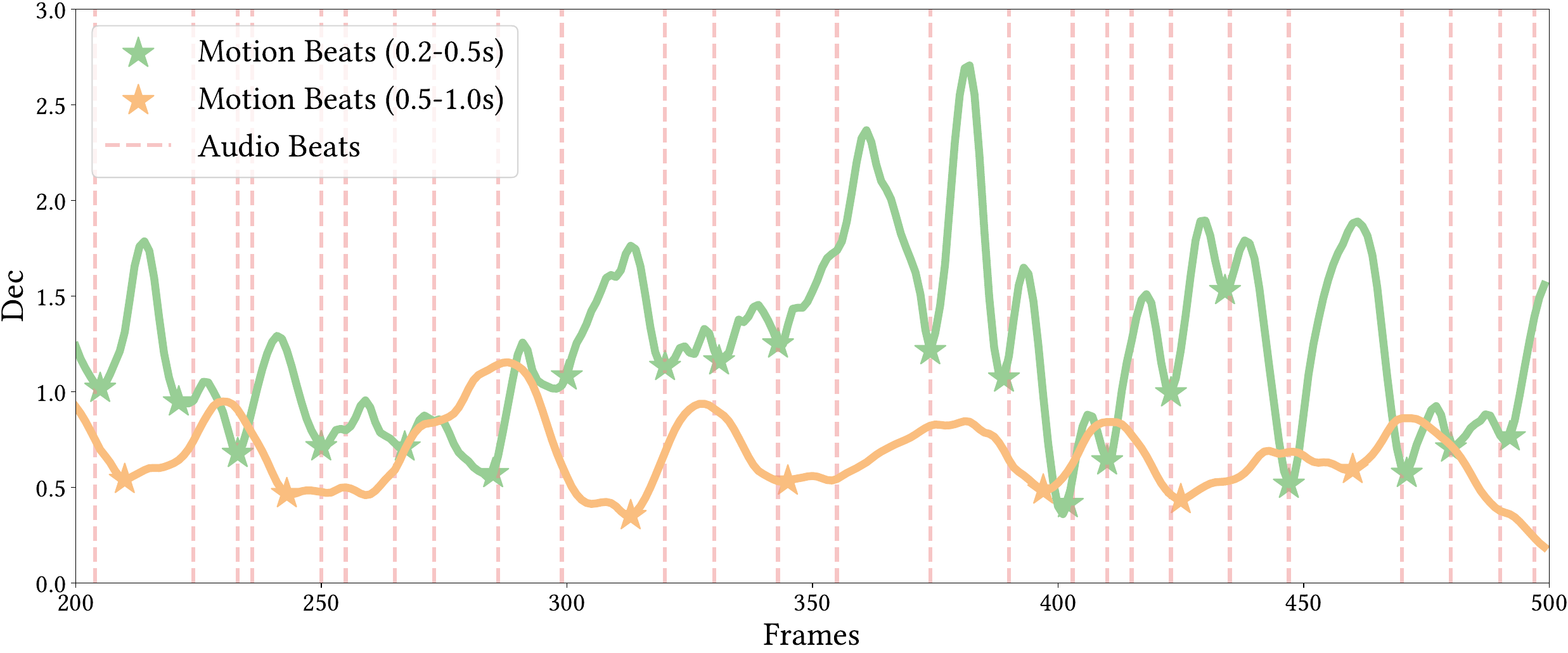}
    \caption{Audio-to-gesture synchronization under different ranges of onset interval. Motion beats are computed based on the local minima of joint deceleration. Audio beats are identified using the audio onsets.}
    \label{fig:fig10}
    \Description{}
\end{figure}

Moreover, to explicitly show the effect of the minimum onset interval, we visualize the synchronization between the motion and audio beats under different interval ranges in \fig\ref{fig:fig10}. 
Similar to \citet{aristidou2021rhythm}, we calculate the motion beats based on the motion deceleration \cite{davis2018visual}.
As shown in Figure \ref{fig:fig10}, the motion beats extracted under the interval range of 0.2-0.5s (denoted by the green stars) synchronize with the audio beats (dashed red lines). In contrast, the motion beats extracted under the interval range of 0.5-1.0s (denoted by the orange stars) do not match well with the audio beats, which proves that a high minimum interval (0.5s) will cause a low beat sensitivity of the model.

\begin{table}[t]
    \centering
    \caption{Effects of the range of onset intervals.}
    \label{tab:table2}
    
    \newcolumntype{Y}{>{\raggedleft\arraybackslash}X}
    \newcolumntype{Z}{>{\centering\arraybackslash}X}
    \begin{tabularx}{\linewidth}{XlYY}
        \toprule
        Dataset & Range of Onset Intervals & FGD $\downarrow$ & PMB ($\%$) $\uparrow$ \\
        \toprule
        \multirow{3}*{Trinity} & 0.5-1.0s & 25.45 & 73.87 \\
        & 0.2-1.0s & 19.16 & 90.75 \\
        & 0.2-0.5s & $\bm{10.78}$ & $\bm{91.36}$ \\
        
        \midrule
        \multirow{3}*{TED} & 0.5-1.0s & 3.61 & 65.34 \\
        & 0.2-1.0s & 2.55 & 89.10 \\
        & 0.2-0.5s & $\bm{2.04}$ & $\bm{89.52}$ \\
        \bottomrule
    \end{tabularx}
    
\end{table}

\begin{figure}[t]
    \centering
    \begin{subfigure}[t]{0.47\linewidth}
        \centering
        \includegraphics[width=\linewidth]{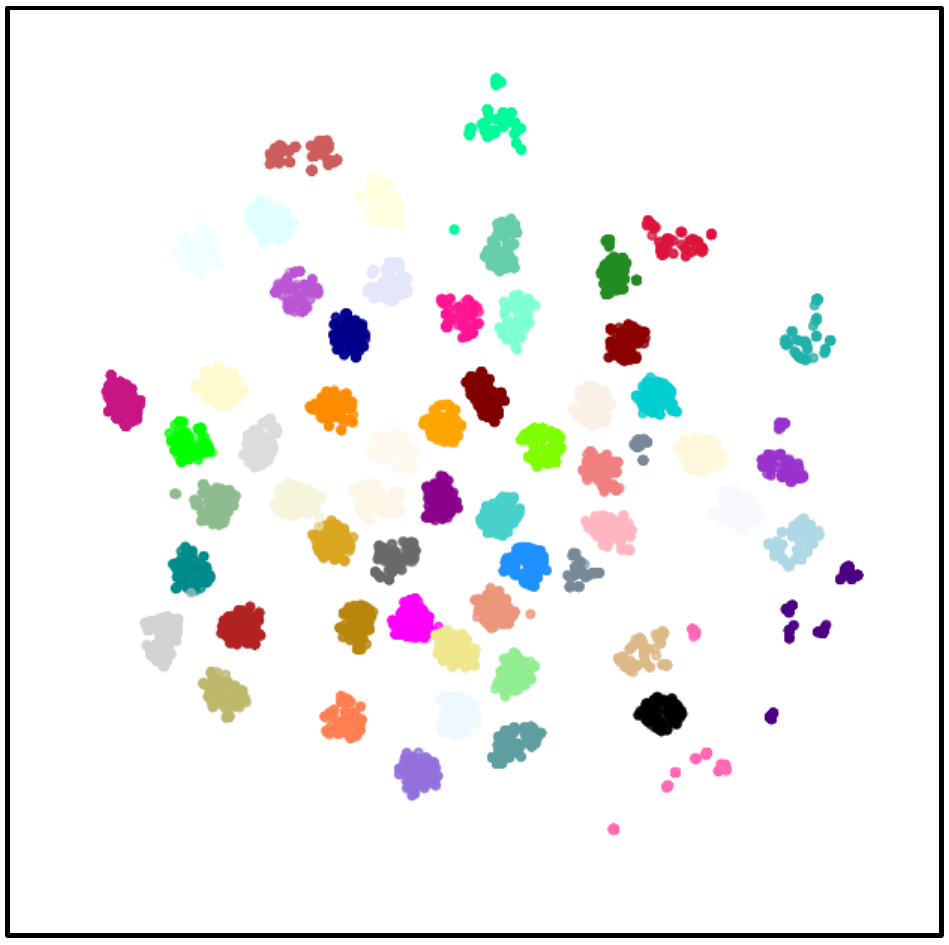}
        \caption{w/ gesture lexeme}
        \label{fig:fig11a}
    \end{subfigure}
    \hspace{\fill}
    \begin{subfigure}[t]{0.47\linewidth}
        \centering
        \includegraphics[width=\linewidth]{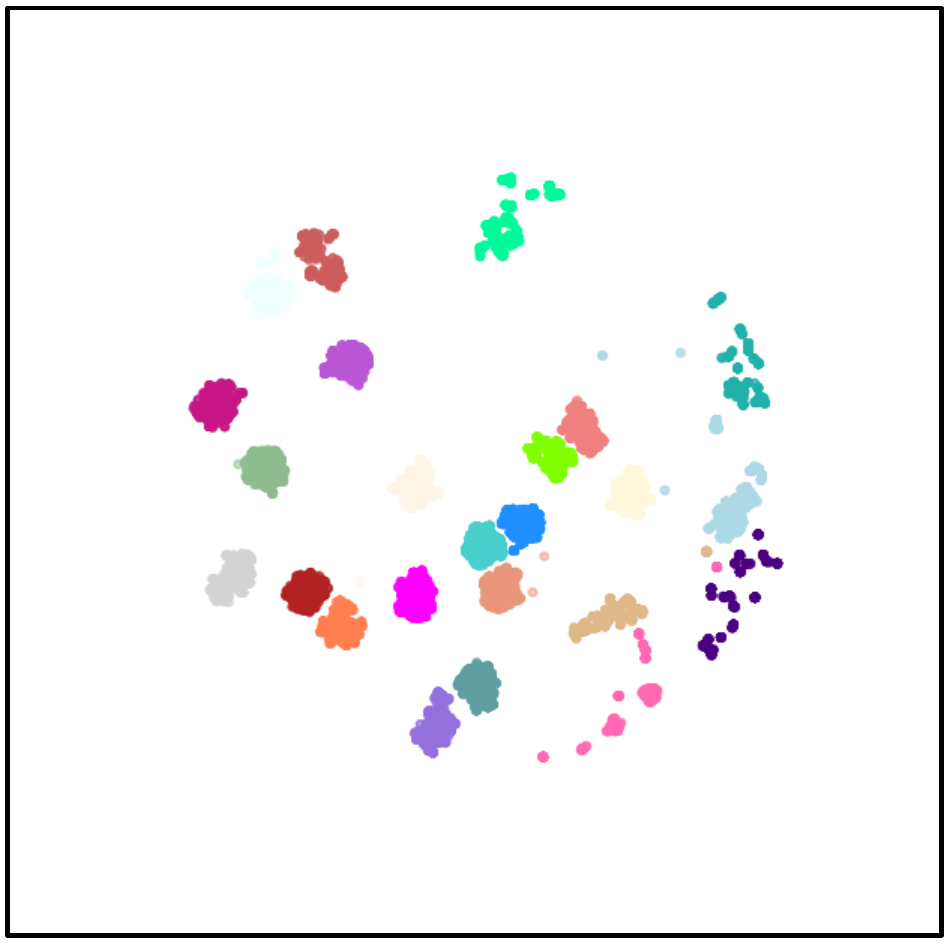}
        \caption{w/o gesture lexeme}
        \label{fig:fig11b}
    \end{subfigure}
    \caption{t-SNE visualization of motion latent codes on the TED dataset, computed using the pre-trained encoder $\mathcal{E}$. Colors represent gesture lexemes.}
    \label{fig:fig11}
    \Description{}
\end{figure}
\begin{table}[t]
    \centering
    \caption{Diversity of motions generated w/ and w/o gesture lexemes.}
    \label{tab:table3}
    
    \newcolumntype{Y}{>{\raggedleft\arraybackslash}X}
    \newcolumntype{Z}{>{\centering\arraybackslash}X}
    \begin{tabularx}{\linewidth}{llY}
        \toprule
        Dataset & System & Diversity $\uparrow$ \\
        \toprule
        \multirow{3}*{Trinity} & Real Gesture & 3.79 \\
        \cline{2-3}
        & w/o gesture lexeme & 1.91 \\
        & w/ gesture lexeme & $\bm{3.40}$ \\
        
        \midrule
        \multirow{3}*{TED} & Real Gesture & 4.32 \\
        \cline{2-3}
        & w/o gesture lexeme & 2.99 \\
        & w/ gesture lexeme & $\bm{4.09}$ \\
        \bottomrule
    \end{tabularx}

\end{table}

\subsubsection{Gesture Lexeme}
\label{subsubsec:gesture_lexeme}
Table \ref{tab:table1} has shown that FGD increases significantly without the gesture lexeme, indicating the importance of the gesture lexeme in achieving quality motions. Besides, as demonstrated in \fig\ref{fig:fig11}, the variety of the generated gestures is also significantly reduced without the gesture lexeme. To show this conclusion quantitatively, we calculate the entropy of the gesture lexemes to measure the motion diversity as
\begin{equation}
    \eqword{Diversity} = -\sum_{i=1}^{N_s}p_i\log{p_i},
\end{equation}
where $N_s$ is the size of gesture lexicon, $p_i$ indicates the occurrence frequency of the $i$-th lexeme in the generated gestures. As shown in Table \ref{tab:table3}, our system with the gesture lexeme creates much higher diversity than the system without it, which further testifies the conclusion of \fig\ref{fig:fig11} and again emphasizes the importance of the gesture lexeme.

\subsubsection{Size of Gesture Lexicon}
\begin{figure}[t]
    \centering
    \begin{subfigure}[t]{0.47\linewidth}
        \centering
        \caption*{Trinity Gesture dataset}
        \includegraphics[width=\linewidth]{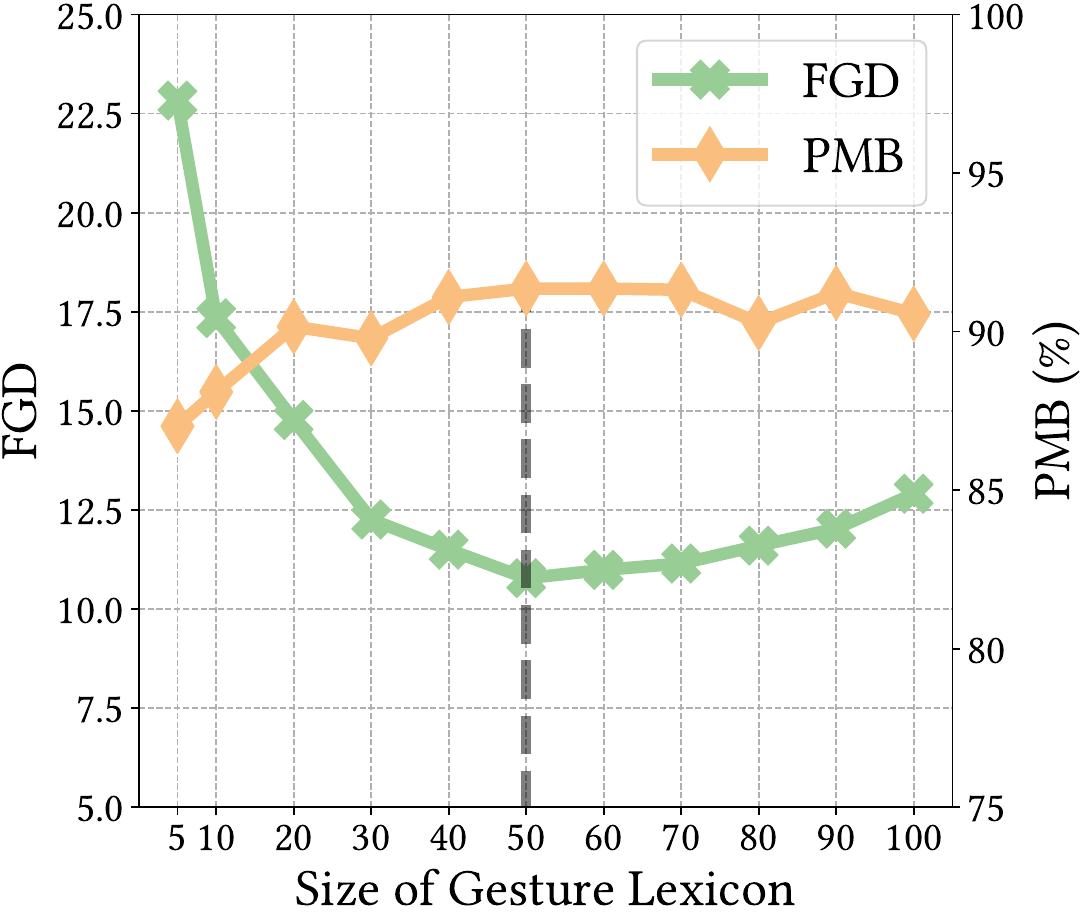}
        \label{fig:fig12a}
    \end{subfigure} 
    \hspace{\fill}
    \begin{subfigure}[t]{0.47\linewidth}
        \centering
        \caption*{TED Gesture dataset}
        \includegraphics[width=\linewidth]{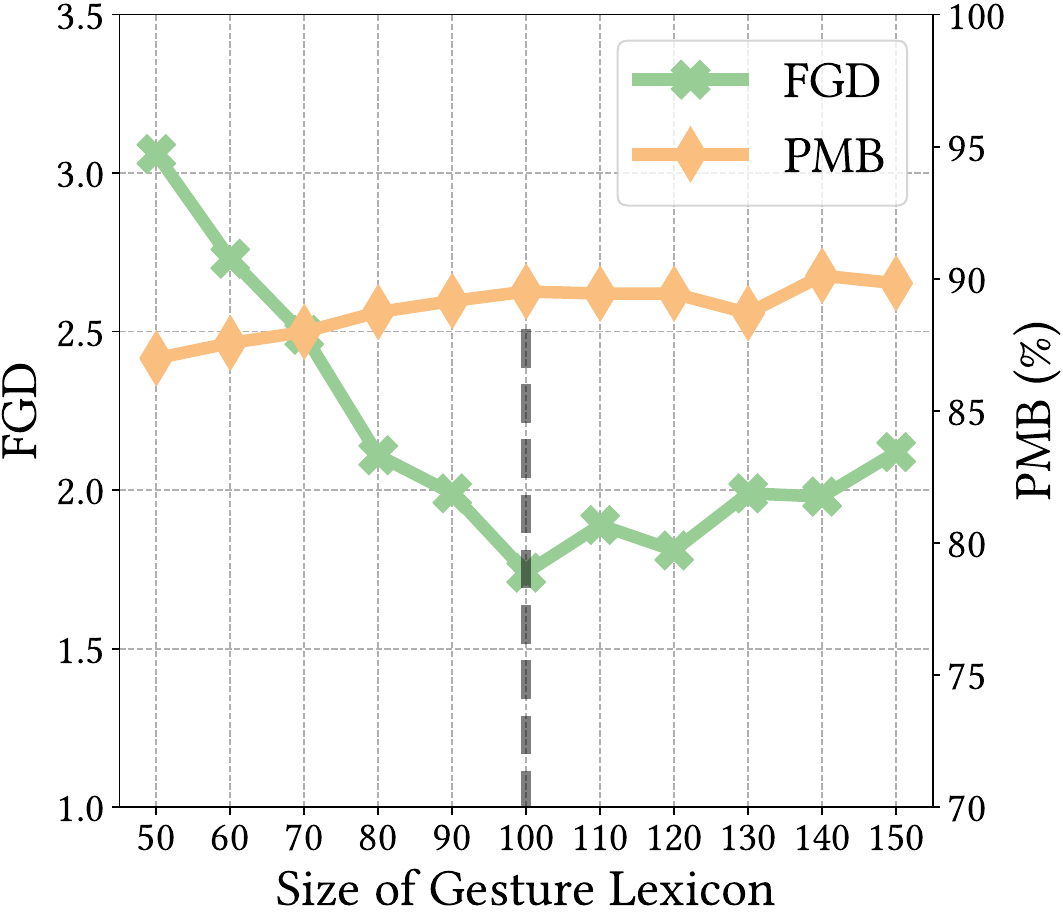}
        \label{fig:fig12b}
    \end{subfigure} 
    \caption{Effects of the size of the gesture lexicon. The gray dashed lines mark the optimal lexicon size.}
    \label{fig:fig12}
    \Description{}
\end{figure}

Figure \ref{fig:fig12} shows the performance of our system under different gesture lexicon sizes. It can be seen that neither too small nor too large lexicons achieve good results, measured as FGD values. On the one hand, a small gesture lexicon forces a diverse range of gesture motions to be merged into the same gesture lexeme, which aggravates the one-to-many mapping issue and causes the generator hard to learn all the motions. On the other hand, an excessively large lexicon forcibly splits many lexemes into sub-lexemes. These sub-lexemes are typically close together, making the gesture lexeme interpretation more challenging and thus negatively affecting the gesture quality. We note that the PMB metric is less affected by the gesture lexicon size, possibly because our beat-based segmentation and normalization mechanism explicitly enforces the gesture rhythm. Based on these experiments, we set the sizes of the gesture lexicon for the Trinity and TED dataset to 50 and 100, respectively, as shown in Figure \ref{fig:gesture_lexicon}.

\subsubsection{Different Interpreters}
\begin{table}[t]
    \centering
    \caption{Comparison of interpreters. The statistical interpreter selects gesture lexemes based on the frequency distribution of lexemes but neglects the input speech. The learning-based interpreter is our default interpreter that translates the input speech into gesture lexemes.}
    \label{tab:table4}
    
    \newcolumntype{Y}{>{\raggedleft\arraybackslash}X}
    \newcolumntype{Z}{>{\centering\arraybackslash}X}
    \begin{tabularx}{\linewidth}{XcYYY}
        \toprule
        \multirow{2}*{Dataset} & Interpreter  & Accuracy  & \multirow{2}*{FGD $\downarrow$} & PMB  \\
        & Type &($\%$)         $\uparrow$ & & ($\%$) $\uparrow$\\
        \toprule
        \multirow{2}*{Trinity} & Statistical & 28.01 & 21.53 & 90.75 \\
        & Learning-based & $\bm{59.15}$ & $\bm{10.78}$ & $\bm{91.36}$ \\
        
        \midrule
        \multirow{2}*{TED} & Statistical & 32.11 & 3.57 & 87.99 \\
        & Learning-based & $\bm{62.53}$ & $\bm{2.04}$ & $\bm{89.52}$ \\
        \bottomrule
    \end{tabularx}

\end{table}

In our system, the lexeme interpreter determines the sequence of gesture lexemes. Our learning-based interpreter described in Section~\ref{subsec:lexeme_interpreter} accomplishes this task according to the speech features and the previous lexemes. Besides, inspired by \cite{aristidou2021rhythm}, we further experiment with a statistical interpreter that matches the frequency of each lexeme in the generated gestures with the reference. Specifically, for a speaker~$\vect{I}$, we calculate the frequency distribution of gesture lexemes, $\vect{f}^{\vect{I}}\in\mathbb{R}^{N_s}$, and a transition matrix, $\vect{L}\in\mathbb{R}^{N_s\times{}N_s}$, describing the frequency of transitions between lexemes using the training data. These quantities can be considered as a global representation of speaker's gesture style. 
During inference, we configure the statistical interpreter to ensure that the lexeme distribution of the generated gesture sequence $\vect{f}_t$ matches the corresponding global distribution $\vect{f}^{\vect{I}}$. 
To achieve this goal, at each generation step, we compute a multinomial distribution characterized by $\vect{f}^*_{t+1}$ using the transition matrix and the difference between the current and target lexeme frequencies $\vect{f}_t$ and $\vect{f}_{\vect{I}}$, where
\begin{equation}
    \vect{f}^*_{t+1} = \softmax(\vect{f}^{\vect{I}}-\vect{f}_t )
    \cdot \vect{L}.
\end{equation}
Then the next lexeme is sampled from the multinomial distribution.
This statistical interpreter does not consider the input speech but only the statistics of the generated gestures when selecting the lexemes. In practice, the result motions still look acceptable but more random. This can be confirmed by Table \ref{tab:table4}, where the statistical interpreter exhibits a lower prediction accuracy and higher FGD values than our learning-based method because of the lack of semantic information brought by the input speech.

\subsubsection{Positional Encoding}
\label{subsubsec:pos_enc}
\begin{table}[t]
    \centering
    \caption{Effects of the positional encoding block.}
    \label{tab:pos_enc}
    
    \newcolumntype{Y}{>{\raggedleft\arraybackslash}X}
    \newcolumntype{Z}{>{\centering\arraybackslash}X}
    \begin{tabularx}{\linewidth}{XlYY}
        \toprule
        Dataset & System & FGD $\downarrow$ & PMB ($\%$) $\uparrow$ \\
        \toprule
        \multirow{2}*{Trinity} & w/o positional encoding & 11.15 & 89.98 \\
        & w/ positional encoding & $\bm{10.78}$ & $\bm{91.36}$ \\
        
        \midrule
        \multirow{2}*{TED} & w/o positional encoding & 2.19 & 88.13 \\
        & w/ positional encoding & $\bm{2.04}$ & $\bm{89.52}$ \\
        \bottomrule
    \end{tabularx}
\end{table}
The positional encoding block (\eqn\ref{eqn:pos_enc}) informs the generator about the frame-level progress of the synthesis in a motion block, which helps the generator model the temporal structure, especially the rhythm, of the sequence. As shown in Table \ref{tab:pos_enc}, the positional encoding block can improve the beat-matched rate (higher PMB) while enhancing the perceptual quality of generated movements (lower FGD).
\section{Conclusion}
\label{sec:conclusion}
In this paper, we present a rhythm- and semantics-aware co-speech gesture synthesis system that can generate realistic gestures to accompany a speech.  For the rhythm, we utilize a segmentation pipeline that explicitly enforces beat alignment to ensure the temporal coherence between the speech and gestures. For the semantics, we successfully disentangle both low- and high-level neural embeddings of speech and motion based on linguistic theory. Then, we devise two neural interpreters to build correspondence between the hierarchical embeddings of the speech and the motion. To evaluate the rhythmic performance, we propose a new objective metric, PMB, to measure the percentage of matched beats. Our method outperforms state-of-the-art systems both objectively and subjectively, as indicated by the MAJE, MAD, FGD, PMB metrics, and human feedback. The cross-language synthesis experiment demonstrates the robustness of our system for rhythmic perception. In terms of application, We show our system's flexible and effective style editing ability that allows editing of several directorial styles of the generated gestures without manual annotation of the data. Lastly, we have systematically conducted detailed ablation studies that justify the design choices of our system.

There is still room for improvement in our current research. 
First, our beat detection algorithm is not perfect. We have assumed that the gesture beats coincide with the verbal stresses, but in practice, it has been observed that gesture beats may not always correspond to stressed syllables \cite{mcclave1994gestural}. 
How to accurately model the complex gestural rhythm is an exciting topic for further exploration.
Second, our system can only capture semantics-related gestures repeatedly appearing in the dataset. Learning semantically meaningful gestures that are sparsely distributed in a dataset and allowing a user to control the gesture corresponding to specific semantics is still challenging.
Third, our system hypothesizes that each audio onset should correspond to a beat gesture. However, in reality, humans do not make a beat gesture at every point of verbal emphasis. We believe our framework can be easily augmented by employing another model to predict whether the character should gesture at a specific moment, as suggested by [Speech2Properties2Gestures], and replacing the corresponding lexeme with the silent lexeme.
Finally, we only consider the upper body gestures in this work. Generating full-body gestures that include locomotion, facial expressions, finger motions, and the temporal and semantic correspondence among them is a valuable topic for future exploration.

%%
%% Acknowledgments.
\begin{acks}
    We thank the anonymous reviewers for their constructive comments.
    This work was supported in part by NSFC Projects of International Cooperation and Exchanges (62161146002).
\end{acks}

\bibliographystyle{ACM-Reference-Format}
\bibliography{gesture}

\end{document}